\documentclass[aip,jcp,preprint,letterpaper,onecolumn]{revtex4-1}
\usepackage[utf8]{inputenc}
\usepackage{modiagram}
\usepackage{caption}
\usepackage{subcaption}
\usepackage{natbib}
\usepackage{bm}
\usepackage{amsmath}
\newcommand{\bra}[1]{\left\langle #1 \right|}
\newcommand{\ket}[1]{\left|#1\right\rangle}
\newcommand{\braket}[2]{\left\langle#1 |  #2\right\rangle}

\newcommand{\expt}[1]{\left\langle#1\right\rangle}

\begin{document}
\title
{Beyond the Coulson-Fischer point: Characterizing single excitation CI and TDDFT for excited states in single bond dissociations}
	\author{Diptarka Hait}
	%\email{diptarka@berkeley.edu}
	%
	\thanks{These authors contributed equally to this work.}
	\affiliation
	{{Kenneth S. Pitzer Center for Theoretical Chemistry, Department of Chemistry, University of California, Berkeley, California 94720, USA}}
	\author{Adam Rettig}
	%\email{adam\_rettig@berkeley.edu}
	\thanks{These authors contributed equally to this work.}
	\affiliation
	{{Kenneth S. Pitzer Center for Theoretical Chemistry, Department of Chemistry, University of California, Berkeley, California 94720, USA}}
	\author{Martin Head-Gordon}
	\email{mhg@cchem.berkeley.edu}
	\affiliation
	{{Kenneth S. Pitzer Center for Theoretical Chemistry, Department of Chemistry, University of California, Berkeley, California 94720, USA}}
	\affiliation{Chemical Sciences Division, Lawrence Berkeley National Laboratory, Berkeley, California 94720, USA}

\begin{abstract}
Linear response time dependent density functional theory (TDDFT), which builds upon configuration interaction singles (CIS) and TD-Hartree-Fock (TDHF), is the most widely used class of excited state quantum chemistry methods and is often employed to study photochemical processes. This paper studies the behavior of the resulting excited state potential energy surfaces beyond the Coulson-Fisher (CF) point in single bond dissociations, when the optimal reference determinant is spin-polarized. Many excited states exhibit sharp kinks at the CF point, and connect to different dissociation limits via a zone of unphysical concave curvature. In particular, the unrestricted M$_S=0$
lowest triplet T$_1$ state changes character, and does not dissociate into ground state fragments. The unrestricted $M_S=\pm 1$ T$_1$ CIS states better approximate the physical dissociation limit, but their degeneracy is broken beyond the CF point for most single bond dissociations. On the other hand, the $M_S=\pm 1$ T$_1$ TDHF states reach the asymptote too soon, by merging with the ground state from the CF point onwards.  Use of local exchange-correlation functionals causes  $M_S=\pm 1$ T$_1$ TDDFT states to resemble their unphysical $M_S= 0$ counterpart. The 2 orbital, 2-electron model system of minimal basis H$_2$ is analytically treated to understand the origin of these issues, revealing that the lack of double excitations is at the root of these remarkable observations. The behavior of excited state surfaces is also numerically examined for species like H$_2$, NH$_3$, C$_2$H$_6$ and LiH in extended basis sets.
\end{abstract}

\maketitle

\section{Introduction} \label{introduction}
Linear-response time-dependent density functional theory\cite{runge1984density,casida1995time,marques2004time,dreuw2005single} (LR-TDDFT) is the most widely used excited state technique at present.  The popularity of LR-TDDFT (henceforth simply referred to as TDDFT) is entirely a consequence of its computational affordability ($O(N^{2-3})$ cost versus molecule size\cite{dreuw2005single}), which permits application to very large systems of hundreds of atoms.\cite{isborn2011excited} Such species are well beyond the reach of more accurate wave function theory approaches like equation of motion coupled cluster\cite{stanton1993equation,krylov2008equation}, or complete active space self-consistent field\cite{roos1980complete} (CASSCF) combined with corrections that include dynamic correlation\cite{andersson1990second,andersson1992second}. At the same time, TDDFT is considerably more accurate than the corresponding Hartree-Fock (HF) based wavefunction methods: specifically single excitation CI (CIS)\cite{foresman1992toward} and time-dependent HF (TDHF)\cite{dirac1930note}, which neglect dynamic correlation entirely.

In practice, TDDFT is plagued with many potential sources of error, despite having the potential to be formally exact\cite{runge1984density} like ground state DFT\cite{hohenberg1964inhomogeneous}. TDDFT errors can roughly be viewed to originate from two sources: failure of the widely used adiabatic local density approximation\cite{marques2004time,dreuw2005single} (ALDA) and errors in the ground state DFT functional. The former generates large errors whenever the targeted state has large doubles (or higher order) character\cite{maitra2004double,levine2006conical}, but is not expected to be a major problem for (almost) purely single excitations\cite{dreuw2005single}. The latter remains a challenge despite the great accuracy of modern ground state density functionals\cite{mardirossian2017thirty,goerigk2017look,hait2018accurate,hait2018accuratepolar}, as TDDFT tends to dramatically augment relatively small ground state failures. The resulting excited state predictions are therefore considerably less reliable than the corresponding ground state calculations. 

The most well known TDDFT failure is the systematic underestimation of excitation energies for charge-transfer (CT) and Rydberg states, on account of delocalization error\cite{perdew1982density,mori2006many,hait2018delocalization} in the underlying functional\cite{dreuw2003long,dreuw2005single}. While delocalization error is an issue for ground state DFT as well, local exchange-correlation functionals predict a particularly poor description of long ranged particle-hole interaction\cite{dreuw2003long,dreuw2005single} in the linear response limit. This becomes an issue whenever an electron undergoes a large spatial shift on account of the excitation, as is the case for CT or Rydberg states. The systematic underestimation (often on the order of 1-2 eV) can be mitigated (and sometimes over-corrected) via use of range separated hybrid functionals with considerable amounts of nonlocal exchange \cite{iikura2001long,tawada2004long,peach2008excitation, sun2015reliable,hait2016prediction}. By contrast, despite, or rather because of being free of delocalization error, CIS is known to yield CT excitations that are too high, a result of lack of orbital relaxation.\cite{subotnik2011communication}

Another well-known failure of TDDFT stems from instability of the ground state Kohn-Sham DFT (KS-DFT)\cite{kohn1965self} solutions against mixing of occupied and virtual orbitals, which is often induced by static correlation. Similarities between the matrix diagonalized to obtain TDDFT excitation energies and the Hessian of the electronic energy against occupied-virtual mixing\cite{thouless1960stability,seeger1977self, bauernschmitt1996stability,bauernschmitt1996treatment} ensures that a negative eigenvalue in the latter (indicating an unstable ground state solution) often leads to a negative or imaginary ``excitation" energy prediction by the former\cite{dreuw2005single}. A stable KS-DFT solution should not lead to such behavior, but sometimes ground state stability can only be achieved via artificial breakdown of spatial or spin symmetry.

A classic example is the breakdown of spin symmetry in unrestricted KS (UKS) and UHF calculations on closed-shell species, when single bonds are stretched beyond a point called the Coulson-Fisher (CF) point\cite{coulson1949xxxiv}. Spin symmetry breaks on account of the lowest triplet (T$_1$) state mixing with the singlet ground (S$_0$) state, and is consequently described as a `triplet instability'\cite{vcivzek1967stability}.  The resulting spin-polarization leads to lower energies overall, but the corresponding asymmetric spin density cannot correspond to a spin-pure wave function. This form of spin contamination in UKS is irrelevant in the dissociation limit as the S$_0$ and T$_1$ are then degenerate. However, it has consequences at shorter separations around the CF point where the fragment spins on the termini of the stretched bond are still partially coupled to each other. Qualitative success of unrestricted Hartree-Fock (UHF) in this regime\cite{szabo2012modern} nonetheless suggests that UKS methods could yield smooth, qualitatively acceptable ground state potential energy surfaces (PES) for single bond dissociation, though some alarming failures by widely used functionals have been reported recently \cite{hait2019wellbehaved}. UKS calculations also guarantee size consistency in the ground state as the system energy at the single bond dissociation limit is identical to sum of energies of isolated fragments. 

It is possible to avoid spin contamination via using only spin-restricted (RKS) orbitals, but this would result in an unstable ground state solution that has an artificially elevated energy relative to the correct dissociation limit of isolated fragments (i.e. on account of artificial contamination from ionic dissociation products). Furthermore, TDDFT calculations on unstable RKS solutions would result in negative or imaginary triplet excitation energies\cite{dreuw2005single}. The lowest TDDFT singlet excitation energy is also known to spuriously go to zero at the dissociation limit for symmetric bonds when RKS orbitals are employed\cite{giesbertz2008failure}. 
Overall, the benefits of spin-polarization for the ground state are well-recognized, and the associated limitations are understood as a usually acceptable price for smoothly joining accurate solutions at equilibrium (restricted) and dissociation (unrestricted spin-polarized). In wave function theory, use of multiple individually optimized HF determinants\cite{gilbert2008self} as a basis for non-orthogonal CI (NOCI)\cite{thom2009hartree} can go a long way towards restoring spin symmetry in both ground and excited states\cite{sundstrom2014non}. A similar approach is possible in DFT\cite{wu2007configuration,kaduk2011constrained}, although the off-diagonal elements are not well-defined.

In this work, we explore the consequences of ground state spin-polarization in DFT (and HF) for excited states computed by TDDFT (and TDHF/CIS). We focus on single bond dissociations, beginning with the toy problem of H$_2$ in a minimum basis for HF and CIS/TDHF. We then move to more realistic basis sets in a variety of stretched single bond systems, including TDDFT as well. A variety of interesting artifacts are found in these results, which have their origins in the neglect of all double excitations, and in the link between characterizing orbital (in)stability and excited states. We suspect that some of these results have been seen by researchers before (some have been mentioned in Ref \onlinecite{casida2017s2}, for instance), but we believe there is no careful study examining this issue in detail. The results indicate that a great deal of caution is needed when using unrestricted orbitals for TDDFT/TDHF/CIS (as well as related methods like CIS(D)\cite{head1994doubles} or CC2\cite{christiansen1995second}) in closed shell systems beyond the CF point!

\section{A concise summary of TDDFT and TDHF}\label{math}
A full derivation of the TDDFT equations involve application of time-dependent external electric fields to ground state KS-DFT solutions and is well described in Ref.  \onlinecite{dreuw2005single}. We therefore only briefly summarize the key results herein. Let the ground state KS determinant have occupied spin orbitals $\phi_{\{i,j,k,l\ldots\}}$ and virtual (unoccupied) spin orbitals $\phi_{\{a,b,c,d\ldots\}}$ from a general exchange-correlation functional $E_{xc}$. $E_{xc}$ can contain orbital dependent terms like HF exchange, as well as purely local contributions from the electron density $\rho(\vec{r})$ alone. Furthermore, let us assume all orbitals are real valued. In the generalized Kohn-Sham framework\cite{seidl1996generalized}, time dependent HF (TDHF) is the special case where $E_{xc}$ contains \textit{only} 100\% exact exchange.

Under these circumstances, the TDDFT excitation energies $\{\omega\}$  are found via solving the following non-Hermitian generalized eigenvalue problem:
\begin{align}
    \begin{pmatrix}
    \mathbf{A} & \mathbf{B}\\
    \mathbf{B} &\mathbf{A}
    \end{pmatrix}\begin{pmatrix}\mathbf{X}\\\mathbf{Y}\end{pmatrix}=\omega     \begin{pmatrix}
   \mathbf{1}&0\\0 & \mathbf{-1}
    \end{pmatrix}\begin{pmatrix}\mathbf{X}\\\mathbf{Y}\end{pmatrix} \label{tddft}
\end{align}
where the $\mathbf{A} $ and $\mathbf{B} $ matrices are:
\begin{align}
 A_{ia,jb}&=\left(\epsilon_a-\epsilon_i\right)\delta_{ij}\delta_{ab}+\braket{ij}{ab}+\bra{ij}f_{xc}\ket{ab}\label{Agen}\\
 B_{ia,jb}&=\braket{ib}{aj}+\bra{ib}f_{xc}\ket{aj}\label{Bgen}
\end{align}
$\epsilon_{p}$ is the energy of orbital $\phi_p$ and the two electron integrals $\braket{ij}{ab}$ and $\bra{ij}f_{xc}\ket{ab}$ are :
\begin{align}
    \braket{ij}{ab}&=\displaystyle\int d\vec{r}\displaystyle\int \phi_{i}\left(\vec{r}\right)\phi_{j}\left(\vec{r'}\right)\dfrac{1}{|\vec{r}-\vec{r'}|}\phi_{a}\left(\vec{r}\right)\phi_{b}\left(\vec{r'}\right)d\vec{r'}\\
       \bra{ij}f_{xc}\ket{ab}&=\displaystyle\int d\vec{r}\displaystyle\int \phi_{i}\left(\vec{r}\right)\phi_{j}\left(\vec{r'}\right)\dfrac{\delta^2 E_{xc}}{\delta \rho\left(\vec{r}\right)\delta \rho\left(\vec{r'}\right)}\phi_{a}\left(\vec{r}\right)\phi_{b}\left(\vec{r'}\right)d\vec{r'}
\end{align}

The corresponding UKS stability  conditions for real valued orbitals are that both $\mathbf{A}+\mathbf{B}$  and $\mathbf{A} -\mathbf{B}$ be positive semidefinite\cite{seeger1977self}. 
The connection between stability and Eqn \ref{tddft} becomes clear when the latter is simplified to:
\begin{align}
    \left(\mathbf{A}-\mathbf{B}\right)\left(\mathbf{A}+\mathbf{B}\right)\left(\mathbf{X}+\mathbf{Y}\right)=\omega^2\left(\mathbf{X}+\mathbf{Y}\right)\label{eqnreform}
\end{align}

$\mathbf{A}$ is typically much larger than $\mathbf{B}$, since the former contains orbital energy differences (that has mean-field one body contributions) that should be much larger than the purely two body terms (which are the sole constituents of $\mathbf{B}$) for KS-DFT to be viable for the ground state\cite{dreuw2005single}. The non-Hermitian nature of Eqn. \ref{tddft} also leads to the possibility of complex eigenvalues if the stability conditions are violated. Partly for this reason, it has been suggested that setting $\mathbf{B}=0$ should be a useful approximation. The resulting eigenvalue equation is simply $\mathbf{AX}=\omega\mathbf{X}$, and is called the Tamm-Dancoff approximation (TDA)\cite{hirata1999time}. This is roughly half as expensive as full TDDFT and has the added benefit of only requiring diagonalization of a Hermitian matrix, precluding any possibility of unphysical imaginary excitation energies. TDDFT and TDA are also formally size-consistent\cite{dreuw2005single}, making them appealing for studying PESs. Indeed, it has been suggested that TDA is arguably more reliable than TDDFT for explorations of PESs\cite{cordova2007troubleshooting}! 

The special case of $E_{xc}$ being purely the HF exchange functional for TDDFT and TDA merits special attention. In this limit, both become pure wave function methods that can exist independent of KS-DFT. TDHF excitation energies in particular are connected to the correlation energy within the random phase approximation (RPA)\cite{dreuw2005single}. TDHF/TDA turns out to be a configuration interaction (CI) method as:
\begin{align}
    A_{ia,jb}&=\bra{\Phi_{i}^a}\mathbf{H}\ket{\Phi_{j}^b}-E_{HF}\label{cis}
\end{align}
i.e. $A_{ia,jb}$ is the matrix element of the molecular Hamiltonian $\mathbf{H}$ between two singly excited determinants $\ket{\Phi_{j}^b}$ and $\ket{\Phi_{i}^a}$, minus the energy $E_{HF}$ of the HF ground state reference $\ket{\Psi}$ . Since $\bra{\Phi_{i}^a}\mathbf{H}\ket{\Phi}=0$ from Brillouin's theorem \cite{szabo2012modern}, we have 
\begin{align}
     \mathbf{H}=\begin{pmatrix}
   E_{HF} &0 \\0 & \mathbf{A}+E_{HF}\mathbf{1}
    \end{pmatrix}
\end{align}
within the Hilbert space spanned by the reference determinant and all single excitations. Consequently, the TDHF/TDA excitation energies are differences between $E_{HF}$ and other eigenvalues of this Hamiltonian, which is exactly configuration interaction with all single (CIS) substitutions from the HF determinant\cite{foresman1992toward}. CIS is therefore a simple and well-behaved member of the TDDFT family of methods. The density overlocalizing tendency of ground state HF however prevents it from attaining quantitative accuracy even for single excitations, leading to errors on the order of 1-2 eV,\cite{dreuw2005single} especially in the form of systematic overestimation for CT excitation energies\cite{dreuw2005single} (in contrast to TDDFT). 

\section{Spin-flipping excitations within TDDFT/TDHF}
Eqn \ref{cis} offers a physical interpretation of the indices $ia$ and $jb$ as representations of excitations from occupied spin orbital $\phi_i$ to virtual spin orbital $\phi_a$ and from occupied spin orbital $\phi_j$ to virtual spin orbital $\phi_b$, respectively. This interpretation can be generalized beyond CIS to TDHF and even to TDDFT, despite the fictitious nature of KS orbitals.

Labeling each occupied-virtual pair by their spins alone (i.e. $\alpha\alpha,\alpha\beta$ etc.) and integrating out spin degrees of freedom in Eqn \ref{Agen}, we find that the structure of the $\mathbf{A}$ matrix is :
\begin{align}
    \mathbf{A}&=\begin{pmatrix}
    \mathbf{A}_{\alpha\alpha,\alpha\alpha} & \mathbf{A}_{\alpha\alpha,\beta\beta}&0&0\\
    \mathbf{A}_{\beta\beta,\alpha\alpha} & \mathbf{A}_{\beta\beta,\beta\beta}&0&0\\
    0&0&\mathbf{A}_{\alpha\beta,\alpha\beta} & 0\\
    0&0&0 & \mathbf{A}_{\beta\alpha,\beta\alpha}\\
    \end{pmatrix}
\end{align}
This leaves 
\begin{align}
\mathbf{A}_{M_S=0}=\begin{pmatrix}
    \mathbf{A}_{\alpha\alpha,\alpha\alpha} & \mathbf{A}_{\alpha\alpha,\beta\beta}\\
    \mathbf{A}_{\beta\beta,\alpha\alpha} & \mathbf{A}_{\beta\beta,\beta\beta}
    \end{pmatrix}
\end{align}    
as the spin-conserving, $M_S=0$ block, since the spin of the occupied electron being excited to a virtual orbital does not change. On the other hand, $\mathbf{A}_{\alpha\beta,\alpha\beta}$ and $\mathbf{A}_{\beta\alpha,\beta\alpha}$ represent spin-flipping $M_S=\mp 1$ blocks, as they depict the transition from an $\alpha$ occupied to a $\beta$ virtual and the reverse, respectively. Similarly,  we find that:
\begin{align}
    \mathbf{B}&=\begin{pmatrix}
    \mathbf{B}_{\alpha\alpha,\alpha\alpha} & \mathbf{B}_{\alpha\alpha,\beta\beta}&0&0\\
    \mathbf{B}_{\beta\beta,\alpha\alpha} & \mathbf{B}_{\beta\beta,\beta\beta}&0&0\\
    0&0&0 & \mathbf{B}_{\alpha\beta,\beta\alpha}\\
    0&0&\mathbf{B}_{\beta\alpha,\alpha\beta} & 0\\
    \end{pmatrix}
\end{align}

It can immediately be seen that the spin-conserving block
\begin{align}
\mathbf{B}_{M_S=0}=\begin{pmatrix}
    \mathbf{B}_{\alpha\alpha,\alpha\alpha} & \mathbf{B}_{\alpha\alpha,\beta\beta}\\
    \mathbf{B}_{\beta\beta,\alpha\alpha} & \mathbf{B}_{\beta\beta,\beta\beta}
    \end{pmatrix}
\end{align}    
is independent of the spin-flipping block, like $\mathbf{A}_{M_S=0}$. Standard TDDFT/TDHF procedures typically focus only on the spin-conserving block, as the eigenvalues obtained from the spin-conserving block alone are a subset of the exact solutions to the full Eqn \ref{eqnreform}.  The spin-flipping block nonetheless does contain physical content, and is essential for obtaining states with different $M_S$ than the reference. For instance, the $M_S=\pm 1$ triplets for a molecule with a singlet ground state can only be obtained from the spin-flipping block, while the $M_S=0$ state can be obtained from the spin-conserving block. It is also worth noting that the spin-flipping blocks of the $\mathbf{A}$ and $\mathbf{B}$ matrices are involved in determining whether UHF solutions are stable against spin-flipping orbital rotations to Generalized HF (GHF) solutions\cite{seeger1977self}. 

Excluding the spin-conserving $M_S=0$ block that perfectly separates from the rest, we have Eqn \ref{tddft} reduce to:
\begin{align}
    \begin{pmatrix}
    \mathbf{A}_{\alpha\beta,\alpha\beta} & 0&0 & \mathbf{B}_{\alpha\beta,\beta\alpha}\\
     0 & \mathbf{A}_{\beta\alpha,\beta\alpha}&\mathbf{B}_{\beta\alpha,\alpha\beta} & 0\\
     0 & \mathbf{B}_{\alpha\beta,\beta\alpha}&\mathbf{A}_{\alpha\beta,\alpha\beta} & 0\\
     \mathbf{B}_{\beta\alpha,\alpha\beta} & 0&0 & \mathbf{A}_{\beta\alpha,\beta\alpha}\\
    \end{pmatrix}\begin{pmatrix}
        \mathbf{X}_{\alpha\beta}\\\mathbf{X}_{\beta\alpha}\\\mathbf{Y}_{\alpha\beta}\\\mathbf{Y}_{\beta\alpha}
    \end{pmatrix}&=\omega \begin{pmatrix}
    \mathbf{1}&0&0&0\\0&\mathbf{1}&0&0\\0&0&-\mathbf{1}&0\\0&0&0&-\mathbf{1}\\
    \end{pmatrix}\begin{pmatrix}
        \mathbf{X}_{\alpha\beta}\\\mathbf{X}_{\beta\alpha}\\\mathbf{Y}_{\alpha\beta}\\\mathbf{Y}_{\beta\alpha}
    \end{pmatrix}\label{sftdhf}
\end{align}
This again can be separated into two independent blocks: 
\begin{align}
        \begin{pmatrix}
     \mathbf{A}_{\beta\alpha,\beta\alpha}&\mathbf{B}_{\beta\alpha,\alpha\beta}\\
     \mathbf{B}_{\alpha\beta,\beta\alpha}&\mathbf{A}_{\alpha\beta,\alpha\beta}\\
    \end{pmatrix}\begin{pmatrix}
        \mathbf{X}_{\beta\alpha}\\\mathbf{Y}_{\alpha\beta}
    \end{pmatrix}&=\omega \begin{pmatrix}
    \mathbf{1}&0\\0&-\mathbf{1}
    \end{pmatrix}\begin{pmatrix}
        \mathbf{X}_{\beta\alpha}\\\mathbf{Y}_{\alpha\beta}
    \end{pmatrix}\label{plus1}
\\
        \begin{pmatrix}
     \mathbf{A}_{\alpha\beta,\alpha\beta}&\mathbf{B}_{\alpha\beta,\beta\alpha}\\
     \mathbf{B}_{\beta\alpha,\alpha\beta}&\mathbf{A}_{\beta\alpha,\beta\alpha}\\
    \end{pmatrix}\begin{pmatrix}
        \mathbf{X}_{\alpha\beta}\\\mathbf{Y}_{\beta\alpha}
    \end{pmatrix}&=\omega \begin{pmatrix}
    \mathbf{1}&0\\0&-\mathbf{1}
    \end{pmatrix}\begin{pmatrix}
        \mathbf{X}_{\alpha\beta}\\\mathbf{Y}_{\beta\alpha}
    \end{pmatrix}\label{minus1}
\end{align}
The indices in Eqn \ref{minus1} can be rearranged to yield:
\begin{align}
    \begin{pmatrix}
     \mathbf{A}_{\beta\alpha,\beta\alpha}&\mathbf{B}_{\beta\alpha,\alpha\beta}\\
     \mathbf{B}_{\alpha\beta,\beta\alpha}&\mathbf{A}_{\alpha\beta,\alpha\beta}\\
    \end{pmatrix}\begin{pmatrix}
      \mathbf{Y}_{\beta\alpha}\\  \mathbf{X}_{\alpha\beta}
    \end{pmatrix}&=\omega \begin{pmatrix}
    -\mathbf{1}&0\\0&\mathbf{1}
    \end{pmatrix}\begin{pmatrix}
      \mathbf{Y}_{\beta\alpha}\\  \mathbf{X}_{\alpha\beta}
    \end{pmatrix}
\end{align}
which is \textit{nearly} identical in structure to Eqn \ref{plus1}, save a sign. It is therefore evident that Eqns \ref{plus1} and \ref{minus1} share the same eigenvectors, and the corresponding eigenvalues differ only by a sign. Although only positive eigenvalues (i.e. excitation energies) have physical meaning, the negative eigenvalues of Eqn \ref{plus1} yield the positive eigenvalues of Eqn \ref{minus1}, and so it suffices to fully solve Eqn \ref{plus1} alone to have all the excitation energies from the spin flipping block. Physically, this can be interpreted as energies for all de-excitations with $M_S=1$ being the negative of excitation energies with $M_S=-1$ (and vice versa).  TDA/CIS here is trivially achieved by diagonalizing the independent $M_S=\pm 1$ blocks $\mathbf{A}_{\beta\alpha,\beta\alpha}$ and $\mathbf{A}_{\alpha\beta,\alpha\beta}$ blocks separately. 

An important difference between the spin-flipped blocks arising from HF and typical KS solutions also merits a mention. To linear response, spin-flipping excitations only affect the off-diagonal $\mathbf{P}_{\alpha\beta}/\mathbf{P}_{\beta\alpha}$ blocks of the one particle density matrix $\mathbf{P}$, and consequently do not affect the electron density at all. Collinear exchange-correlation kernels $f_{xc}$ will therefore have zero contribution from the local component of the exchange-correlation functional. In other words, matrix elements involving $f_{xc}$ will be zero in the spin-flipping block for purely local functionals like PBE\cite{PBE}, and only contributions from HF exchange will count for hybrid functionals like PBE0\cite{pbe0} or  LRC-$\omega$PBEh\cite{lrcwpbeh}.  This means that the local exchange-correlation contribution to Eqn. \ref{Agen}  via orbital energy differences will go uncorrected in KS theory, which (as we will demonstrate) leads to unusual behavior for $M_S=\pm 1$ solutions for TDDFT/TDA relative to CIS. Furthermore, there will be no contributions from local exchange-correlation terms to the $\mathbf{B}$ matrix within the spin-flip block, rendering TDA identical to full TDDFT for local functionals.

\section{CIS and TDHF for stretched H$_2$ in a minimum basis.}

\begin{figure}[bht!]
        %\centering
        \includegraphics[width=0.7\textwidth]{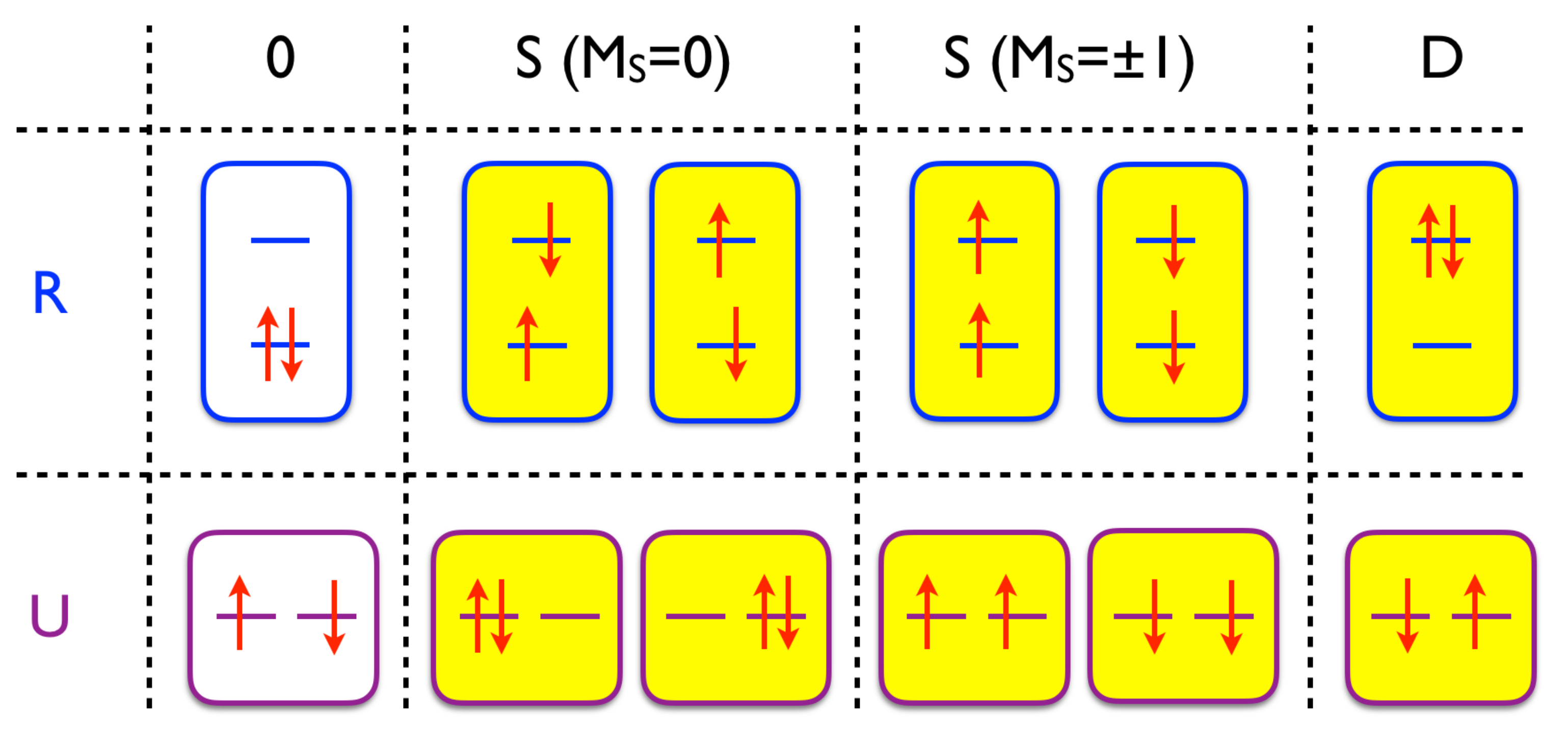}
        \caption{All possible Slater determinants for minimal basis H$_2$ for both restricted (R) and (dissociation-limit) spin-polarized unrestricted (U) solutions to the ground state determinant (0). Aside from the ground state, the four possible single (S) substitutions, and the one double (D) substitution are also depicted.}
        \label{fig:MBS_dets}
\end{figure}

Let us first consider the instructive toy model of minimal basis (STO-3G)\cite{hehre1969self} H$_2$, which contains 2 spatial orbitals and 2 electrons. Six determinants are consequently possible, and are illustrated in Fig. \ref{fig:MBS_dets}. The $M_S=0$ subspace has four determinants and the $M_S=\pm 1$ subspaces have one each. These determinants can be formed from either restricted (R) or unrestricted spin-polarized (U) orbitals. The R orbitals are bonding ($\sigma$) and antibonding ($\sigma^*$) respectively, while the U orbitals at dissociation are atomic orbitals (1s$_{\textrm{A}}$ and 1s$_{\textrm{B}}$). Exact full configuration interaction (FCI) is invariant to the choice of orbitals,  but  HF and TDHF/CIS show critical differences between the R and U cases. 

FCI in the rank-4 $M_S=0$ manifold yields three singlets and a triplet at all possible distances.  At equilibrium, these states are (roughly) a  $\ket{\sigma\bar{\sigma}}$ singlet ground state (X$^1\Sigma_{\rm{g}}$), a $\ket{\sigma^*\bar{\sigma^*}}$ doubly excited singlet (A$^1\Sigma_{\rm{g}}$), along with a singlet ($^1\Sigma_{\rm{u}}$) and a triplet ($^3\Sigma_{\rm{u}}$) resulting from linear combinations of the single excitations $\ket{\sigma\bar{\sigma^*}}$ and $\ket{\sigma^*\bar{\sigma}}$. The two additional states with $M_S=\pm 1$ complete the triplet manifold. In the dissociation limit, the lowest $^1\Sigma_{\rm{g}}$ state and $^3\Sigma_{\rm{u}}$ become degenerate, arising from the four possible ways in which the spins on two isolated H atoms can couple.  There are also two degenerate ($^1\Sigma_{\rm{g}}$ and $^1\Sigma_{\rm{u}}$) higher energy charge transfer (CT) states corresponding to superpositions of $^-$H$\cdots$H$^+$ and $^+$H$\cdots$H$^-$. 

It is instructive to consider the behavior of HF/CIS against this exact behavior.  At internuclear distances smaller than the CF point, the stable HF ground state is $\ket{\sigma\bar{\sigma}}$ (as seen in Fig. \ref{fig:MBS_dets}) and has no spin-polarization. There are 2 possible $M_S=0$ single substitutions ($\ket{\sigma\bar{\sigma^*}}$ and $\ket{\sigma^*\bar{\sigma}}$, as shown in Fig. \ref{fig:MBS_dets}), which arise from $\sigma\to\sigma^*$ transitions. Diagonalizing the CIS Hamiltonian therefore gives a singlet and a triplet state (which are positive and negative linear combinations of $\ket{\sigma\bar{\sigma^*}}$ and $\ket{\sigma^*\bar{\sigma}}$ with equal weights). The $M_S=\pm 1$ states (which are degenerate with the $M_S=0$ triplet) complete the triplet manifold. The situation corresponds very well to FCI (indeed, the $^1\Sigma_{\rm{u}}$ and $^3\Sigma_{\rm{u}}$ levels are exact), save the absence of the doubly excited singlet, which is beyond the scope of a `singles-only' method like CIS.

Let us now consider the (unrestricted) dissociation limit. The stable UHF ground state has an electron localized on each atom (lower panel of Fig. \ref{fig:MBS_dets}), and is spin-contaminated (equal parts singlet and triplet). This spin-polarized UHF state is energetically preferred over RHF due to the absence of spurious ionic (CT) contributions\cite{szabo2012modern} in the former, which make up 50\% of the wave function in the latter. There are two $M_S=0$ single substitutions (Fig. \ref{fig:MBS_dets}) which are singlet with CT character (i.e. H$^-\cdots$H$^+$ and H$^+\cdots$H$^-$). These two singly excited CT determinants are noninteracting, making them UCIS eigenstates which exactly match the FCI CT states.  The remaining singles are the covalent $M_S=\pm 1$ states, which also match FCI. So, at dissociation, UCIS is exact for the $M_S=\pm 1$ components of the FCI triplet, but entirely fails to describe the $M_S=0$ sub-level of the state! A portion of the $M_S=0$ triplet survives via mixing with the ground state to induce spin polarization, while the remainder lies in the omitted doubly excited D determinant.

\begin{figure}[h]
    \centering
    \begin{subfigure}[b]{0.48\textwidth}
        \centering
        \includegraphics[width=\linewidth]{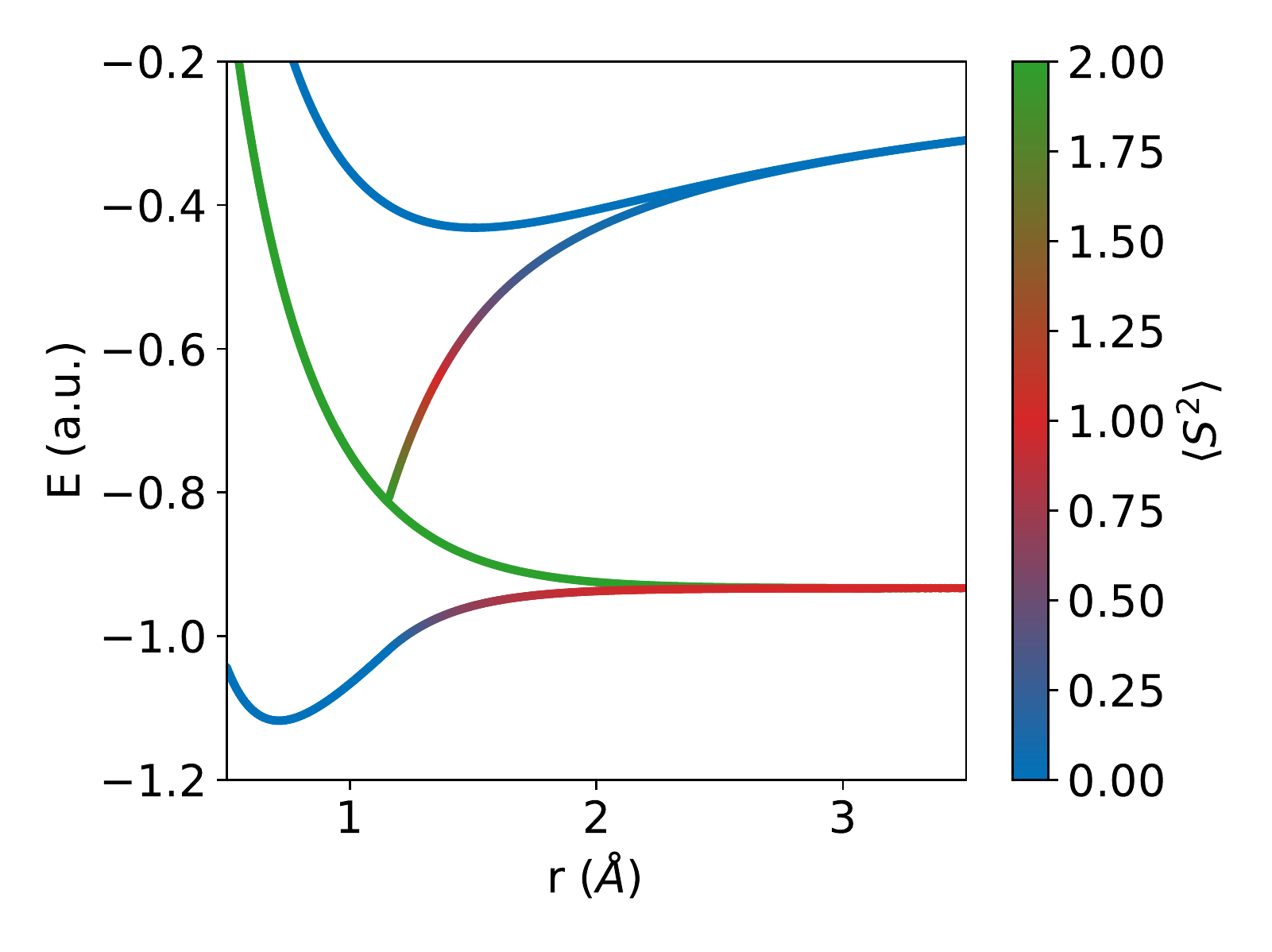}
        \caption{HF and CIS}
        \label{h2min:E}
    \end{subfigure}
    \begin{subfigure}[b]{0.48\textwidth}
        \centering
        \includegraphics[width=\linewidth]{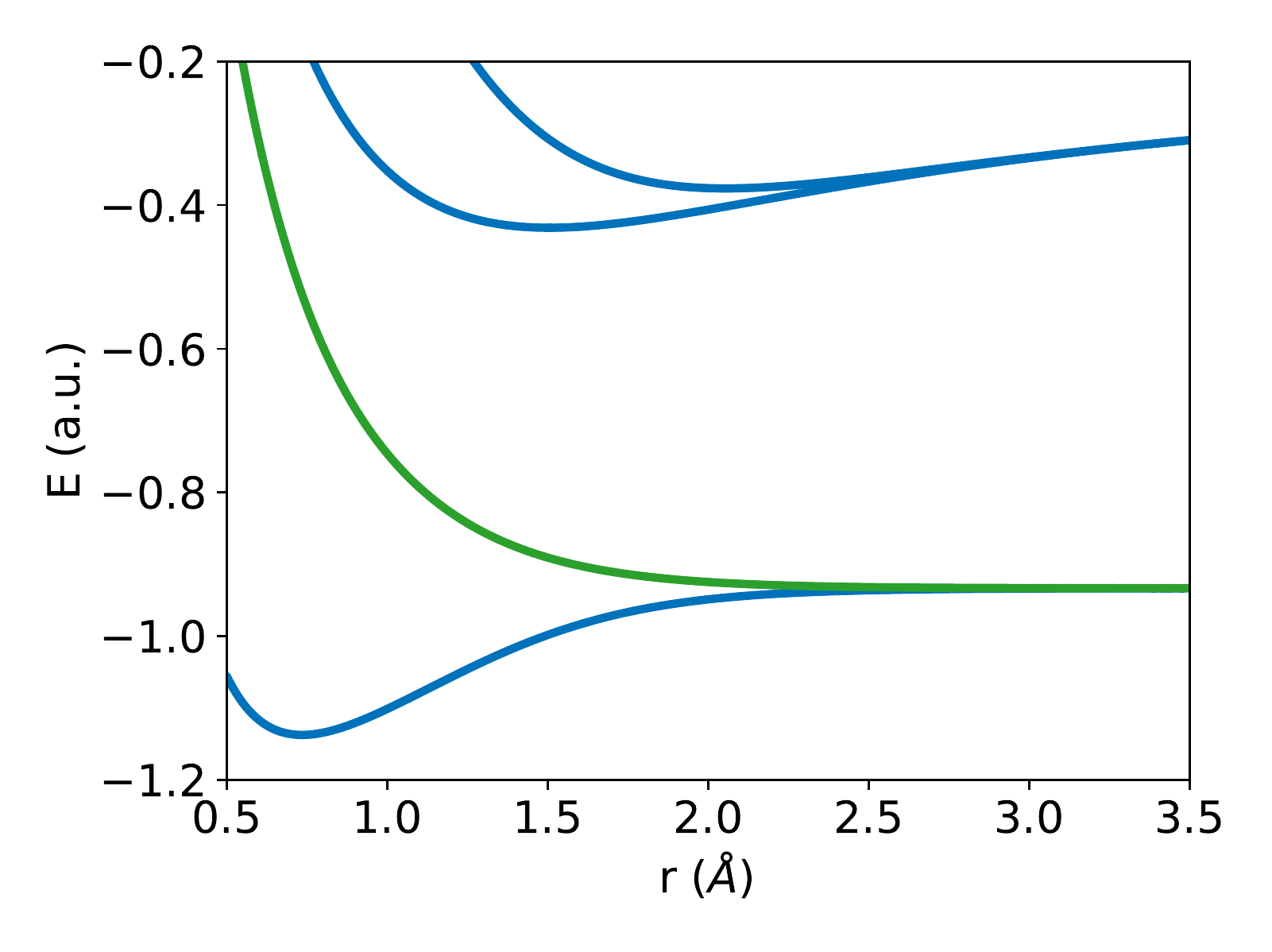}
        \caption{Exact}
        \label{h2min:Eexact}
    \end{subfigure}
    \caption{HF and CIS PESs of H$_2$ in the minimal basis, based on the stable determinant (i.e. RHF for $r<r_{\textrm{CF}}$, and spin polarized UHF for $r \ge r_{\textrm{CF}}$) compared to exact FCI (which is CISD). The $M_S=0$ UCIS T$_1$ state has a pronounced kink at the CF point and subsequently ascends to the CT dissociation limit. However, the $M_S=\pm 1$ T$_1$ states continue on to the proper dissociation limit of neutral atoms in the ground state.}
\end{figure}

The complete FCI PESs are shown in Fig. \ref{h2min:Eexact}, which can be compared against the HF/CIS results in Fig. \ref{h2min:E}. The  FCI T$_1$ surface is nonbonding, and the energy decreases monotonically with internuclear separation $r$. As already discussed above, the RCIS T$_1$ is exact. However, spin polarization in the UHF ground state leads to a very sharp and unphysical kink in the UCIS $M_S=0$ T$_1$ state at the CF point (spuriously suggesting a local minimum), followed by a monotonic rise in energy to the CT dissociation limit, versus the desired neutral atom limit. This state therefore changes character from triplet at the CF point to CT singlet as dissociation approaches, consistent with the analysis given above.   In contrast, the UCIS $M_S=\pm 1$ components of the T$_1$ state remain exact.

Additional insight into the curves can be gained by monitoring spin-polarization in the stable UHF determinant, $\left| {{\Phi _{\textrm{U}}}} \right\rangle  = {2^{ - {1 \mathord{\left/
 {\vphantom {1 2}} \right.
 \kern-\nulldelimiterspace} 2}}}\det \left\{ {\left| A \right\rangle \left| {\bar B} \right\rangle } \right\}$, via a parameter $\theta  \in \left[ {0,\frac{\pi }{4}} \right]$ for orbital mixing. Here $\left| A \right\rangle  = \left| \sigma  \right\rangle \cos \theta  + \left| {{\sigma ^ * }} \right\rangle \sin \theta $ and $\left| B \right\rangle  = \left| \sigma  \right\rangle \cos \theta  - \left| {{\sigma ^ * }} \right\rangle \sin \theta $. Thus $\theta=0$ for $r<r_{\textrm{CF}}$ (RHF regime), and $\theta = \frac{\pi}{4}$ at the dissociation limit (as described in Sec 3.8.7 of Ref \onlinecite{szabo2012modern}). Consequently, ground state $\expt{S^2} = \sin^2 2\theta$, which shows monotonic change from a pure singlet ($\expt{S^2}=0$ for $\theta=0$) at $r=r_{\textrm{CF}}$ to equal singlet-triplet mixture ($\expt{S^2}=1$ for $\theta=\frac{\pi}{4}$) at the dissociation limit. Similarly, the $M_S=0$ UCIS T$_1$ state has $\expt{S^2}=2\cos^22\theta$, and thus begins as a pure triplet ($\expt{S^2}=2$ for $\theta=0$), followed by spin polarization past the CF point to be a singlet-triplet mixture, and ultimately becomes a pure singlet at dissociation ($\expt{S^2}=0$ for $\theta=\frac{\pi}{4}$), showing a complete change of character. In contrast, the spin-polarized continuation of the $M_S=\pm 1$ components of the RCIS triplet remain exact (and smooth) beyond the CF point because they are already exact as single determinants. 

Likewise, the lowest singlet S$_1$ surface is exact, smoothly changing from a valence to CT excited state. It has $\expt{S^2}=0$ for all $\theta$, and therefore has no triplet character. It therefore appears that the continuation of the T$_1$ state is mixing only with the doubly excited S$_2$ state beyond the CF point in the minimal basis picture.

\begin{figure}[h!]
    \begin{subfigure}[b]{0.48\textwidth}
        \centering
        \includegraphics[width=\linewidth]{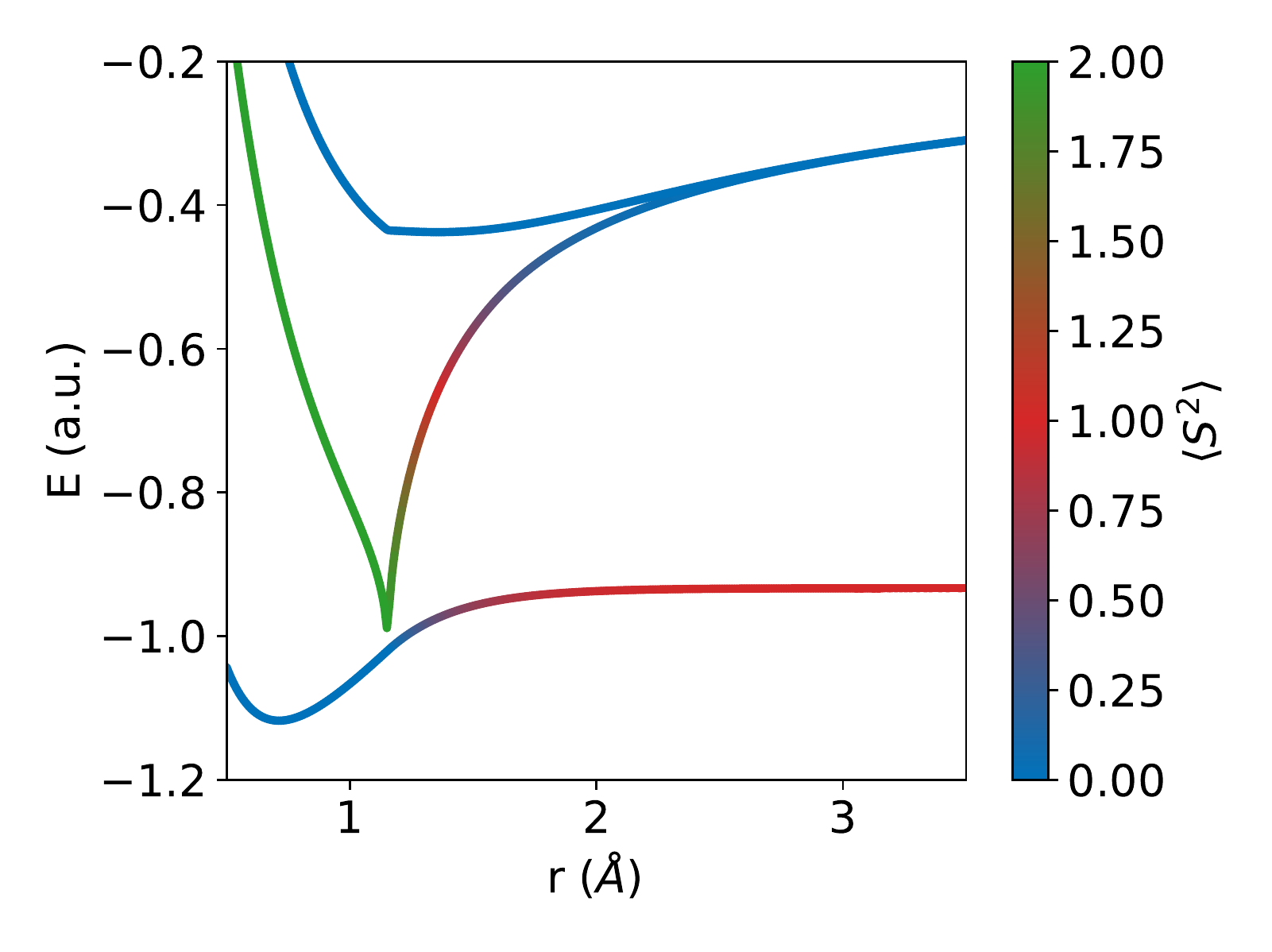}
        \subcaption{$M_S=0$ block}
        \label{h2min:TDHFms0}
    \end{subfigure}
        \begin{subfigure}[b]{0.48\textwidth}
        \centering
        \includegraphics[width=\linewidth]{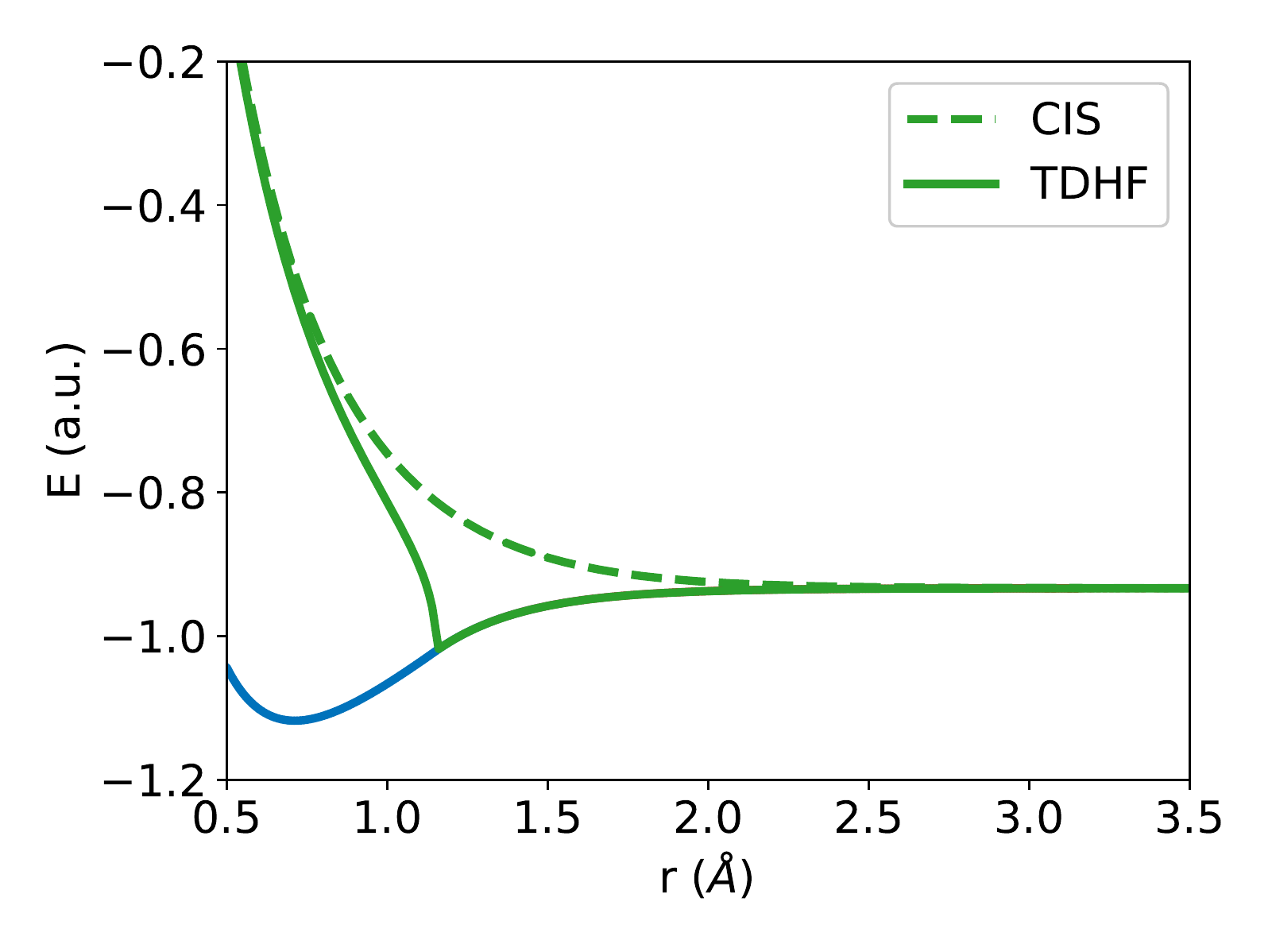}
        \subcaption{Spin-flipping block}
        \label{h2min:TDHFms1}
    \end{subfigure}
    \caption{TDHF excited states for minimal basis H$_2$ ($\expt{S^2}$ values correspond to the corresponding CIS states). The $M_S=\pm 1$ CIS solution is also supplied for comparison.}
   \label{h2min:TDHF}
\end{figure}

It is also instructive to consider the behavior of TDHF proper to see how the inclusion of the $\mathbf{B}$ matrix affects CIS results. Fig \ref{h2min:TDHFms0} reveals that the TDHF states show essentially the same general picture as CIS within the spin-conserving $M_S=0$ block. The kink in the T$_1$ state is even more pronounced, as the surface effectively funnels down to the ground state (i.e. zero excitation energy) at the CF point, before an even more steep ascent to the incorrect dissociation limit. This is a consequence of $\mathbf{A}+\mathbf{B}$ having a zero eigenvalue at the CF point due to onset of spin-polarization induced instability, which leads to a zero eigenvalue for Eqn. \ref{eqnreform}. The S$_1$ state also has a weak kink in the CF point (unlike the case of CIS, where it was exact), but both excited states go to the exact CT dissociation limit. Overall, the performance of full TDHF is somewhat worse than the already poor performance of CIS, consistent with earlier observations \cite{cordova2007troubleshooting}.

The behavior of TDHF in the spin-flip blocks is quite distinct, as made evident by  Fig \ref{h2min:TDHFms1}. The spin-flip T$_1$ states are degenerate with the spin-conserving one prior to spin-polarization, going to the expected zero excitation energy at the CF point and exhibiting a derivative discontinuity therein. They however subsequently remain degenerate with the UHF ground state (i.e. have zero excitation energy). An analytic proof for the zero spin-flip TDHF excitation energy for this toy model is supplied in the Appendix. A more general argument however can be derived from GHF stability theory. The direction of the spin-density induced by the spin-polarization is arbitrary within GHF theory (unlike in UHF where it is constrained to be along the $z$ direction), and therefore orbital rotations that break $\hat{S}_x$ and $\hat{S}_y$ symmetries do not have any associated energy barrier or restoring force\cite{cui2013proper}. Consequently, the GHF stability Hessian\cite{seeger1977self} has two zero eigenvalues corresponding to these orbital rotation normal modes. These lead to four zero eigenvalues in Eqn \ref{sftdhf}\cite{cui2013proper} and two zero eigenvalues in Eqn \ref{plus1}. These $0$ eigenvalues are natural continuations of the pre-CF point T$_1$ curves and consequently can be viewed as T$_1$ excitation energies even beyond the CF point .

Let us briefly summarize this minimal basis H$_2$ story of disaster and success beyond the CF point. The UHF ground S$_0$ state separates correctly with spin-polarization. However, spin-polarization causes the CIS T$_1$ to separate correctly only in the $M_S=\pm 1$ sub-levels, while the $M_S=0$ sub-level becomes a CT singlet at separation. The CIS S$_1$ state however remains exact. TDHF further worsens the CIS reuslts, with the $M_S=0$ T$_1$ solution possessing a more pronounced kink that spuriously connects it to the S$_0$ state. A kink is also induced in the previously perfect S$_1$ state. TDHF within the spin-flip block reveals that while the T$_1$ states remain degenerate with the $M_S=0$ solution until the CF point, they separate afterwards, with the continuations of the spin-flip T$_1$ states becoming degenerate with the UHF ground state post-spin polarization.

\section{Stretched H$_2$ in a larger basis.}
\begin{figure}[htb!]
\vspace*{-15pt}
    \begin{subfigure}[b]{0.48\textwidth}
        \centering
        \includegraphics[width=\linewidth]{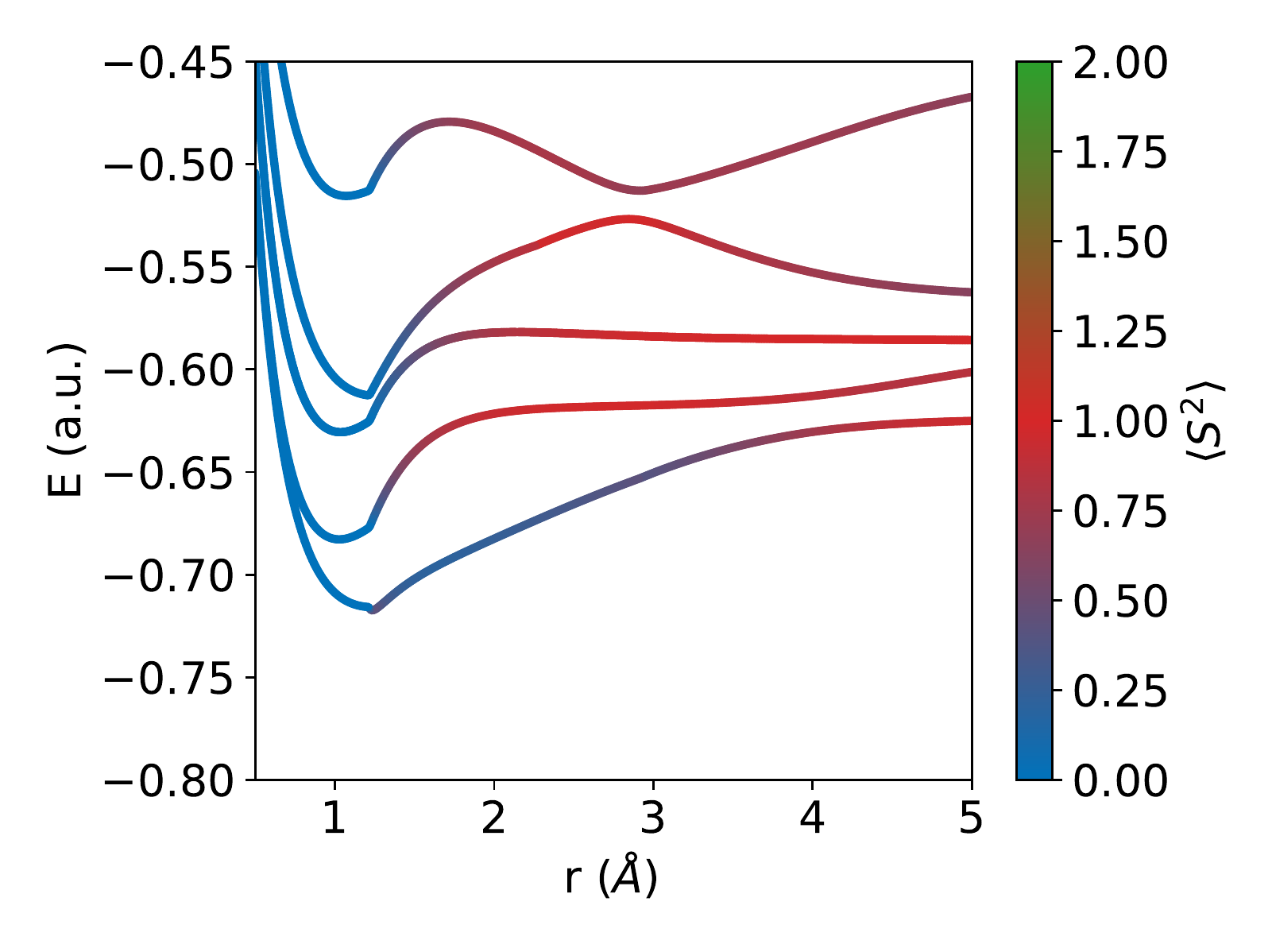}
        \caption{UCIS}
        \label{h2tz:Esing}
    \end{subfigure}
    \begin{subfigure}[b]{0.48\textwidth}
        \centering
        \includegraphics[width=\linewidth]{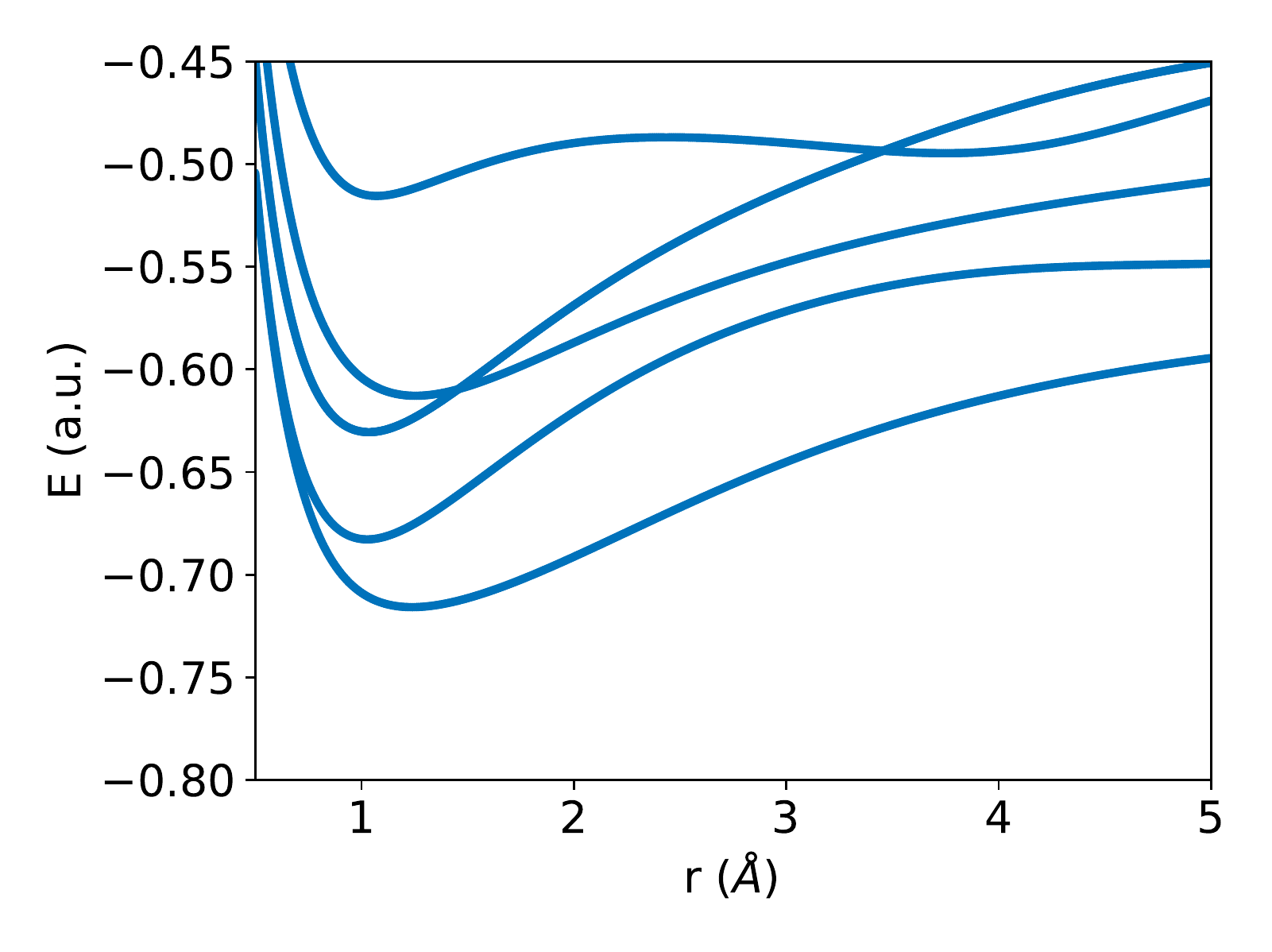}
        \caption{RCIS}
        \label{h2tz:Esingrcis}
    \end{subfigure}
        \begin{subfigure}[b]{0.48\textwidth}
        \centering
        \includegraphics[width=\linewidth]{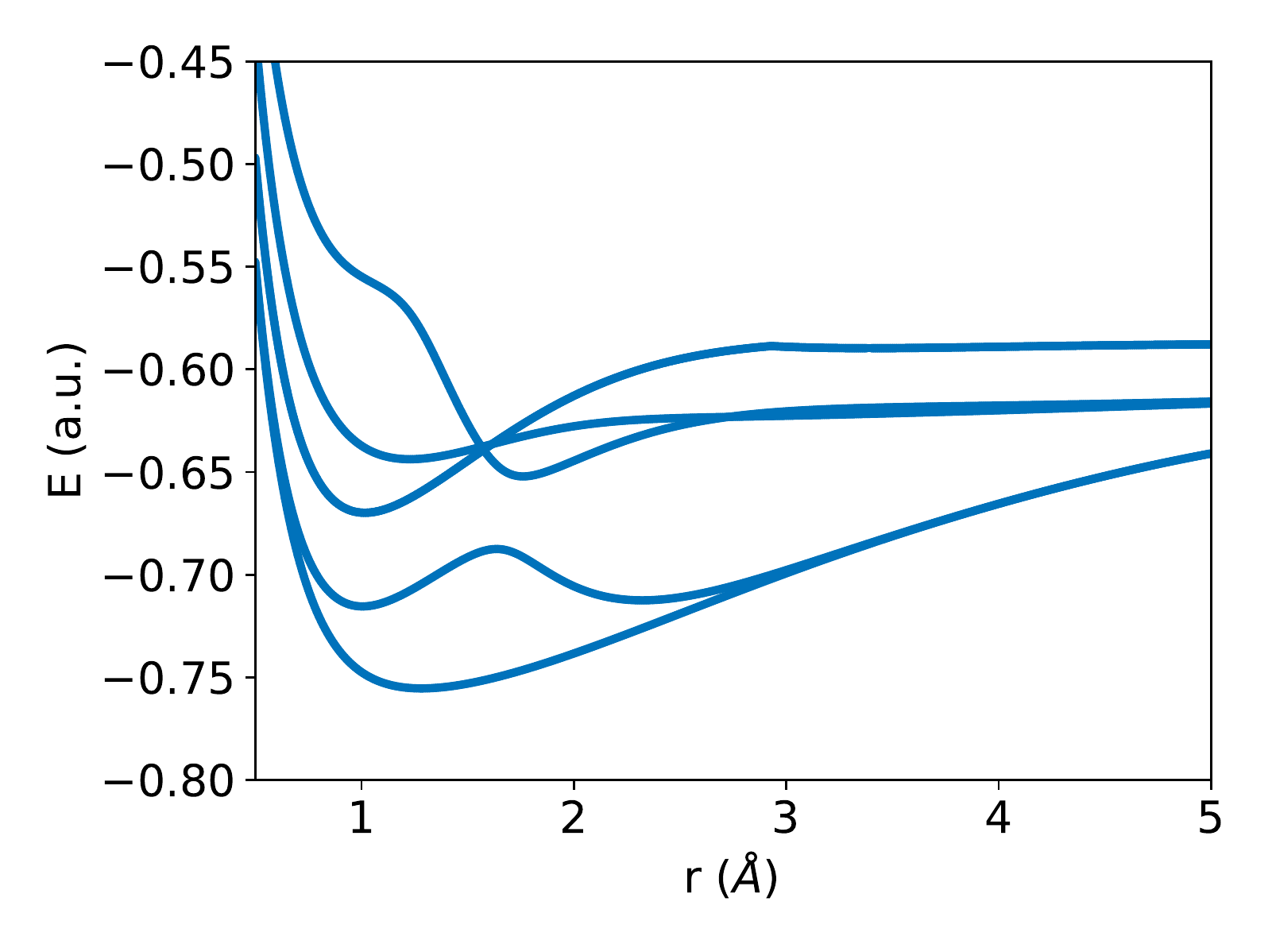}
        \caption{Exact}
        \label{h2tz:Esingexact}
    \end{subfigure}
    \vspace*{-15pt}
    \caption{Low lying CIS singlet excited state PESs of H$_2$ in the aug-cc-pVTZ basis compared to exact results. Small state crossing induced discontinuities might be present on the top surface.  The UCIS singlet states are spin-contaminated beyond the CF point, but their analytical continuation is still followed to the dissociation limit. The S$_0$ ground state has not been depicted, for clarity. }
    \label{h2tz:Sing}
    \vspace*{-15pt}
\end{figure}

\begin{figure}[htb!]
\vspace*{-15pt}
    \begin{subfigure}[b]{0.48\textwidth}
        \centering
        \includegraphics[width=\linewidth]{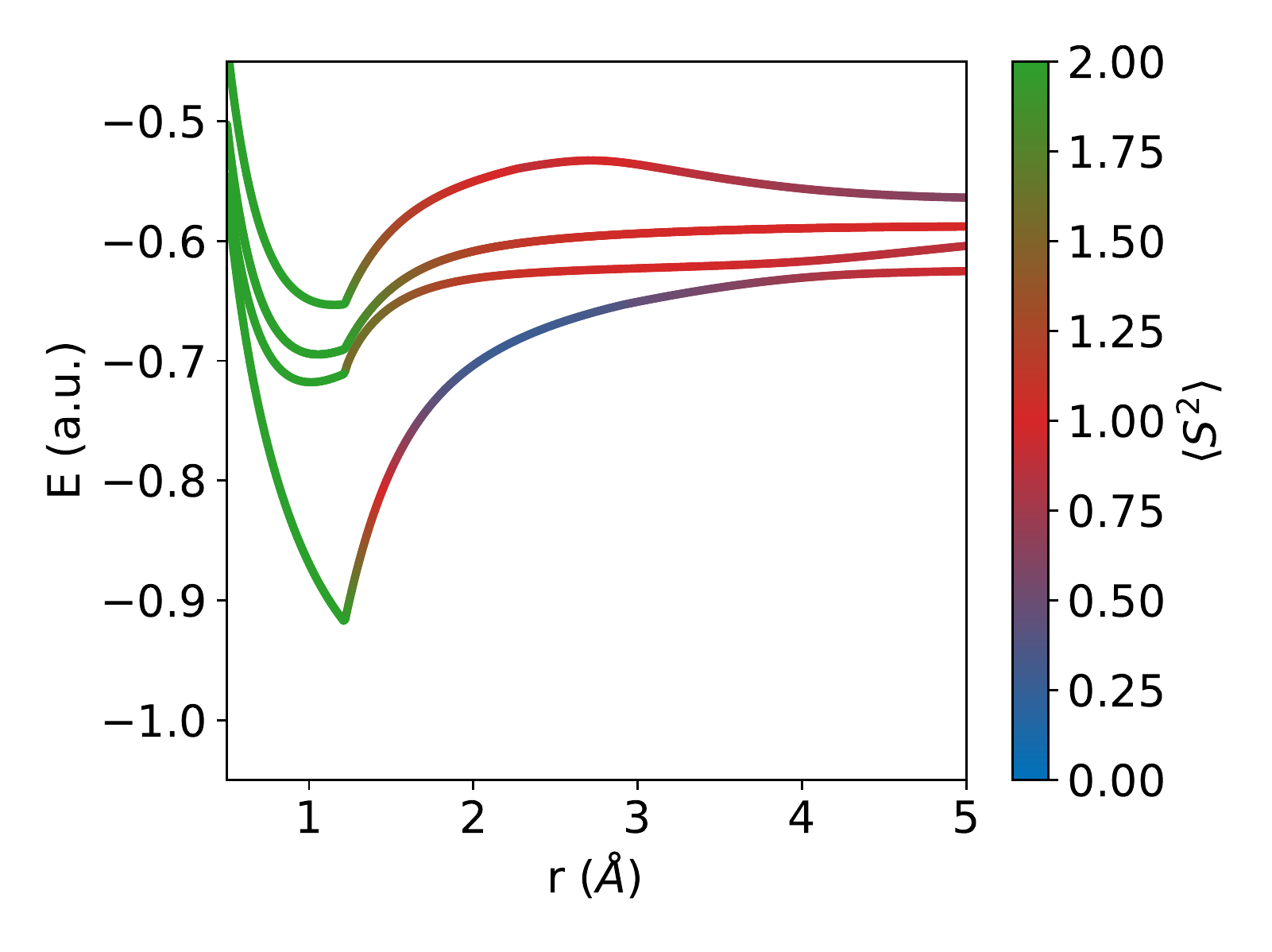}
        \caption{UCIS $M_S=0$.}
        \label{h2tz:Etrip}
    \end{subfigure}
    \begin{subfigure}[b]{0.48\textwidth}
    \centering
    \includegraphics[width=\linewidth]{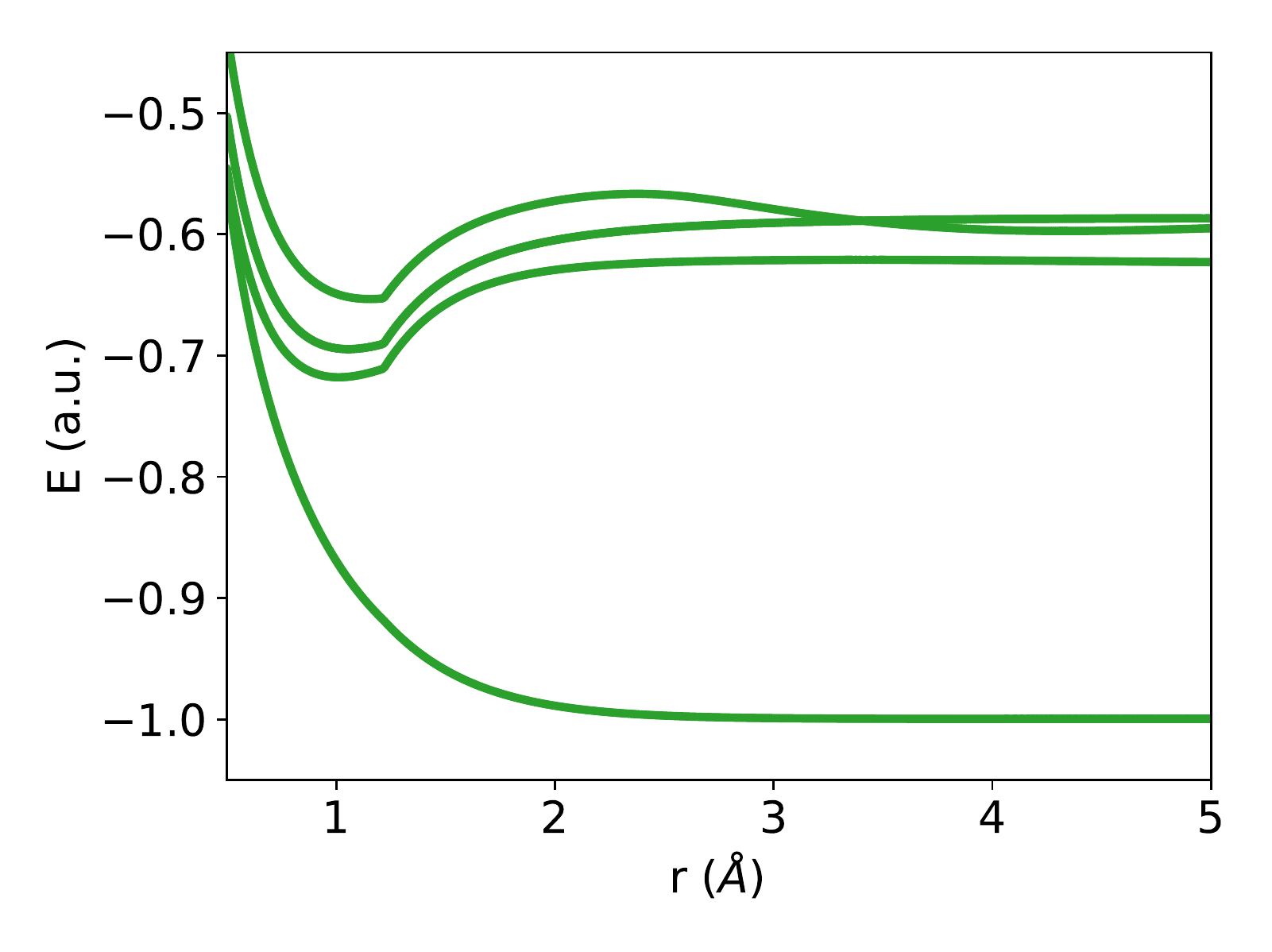}
    \caption{UCIS $M_S=\pm 1$.}
    \label{fig:h2_tz_sfs}
    \end{subfigure}
    \vspace*{-15pt}
    \begin{subfigure}[b]{0.48\textwidth}
        \centering
        \includegraphics[width=\linewidth]{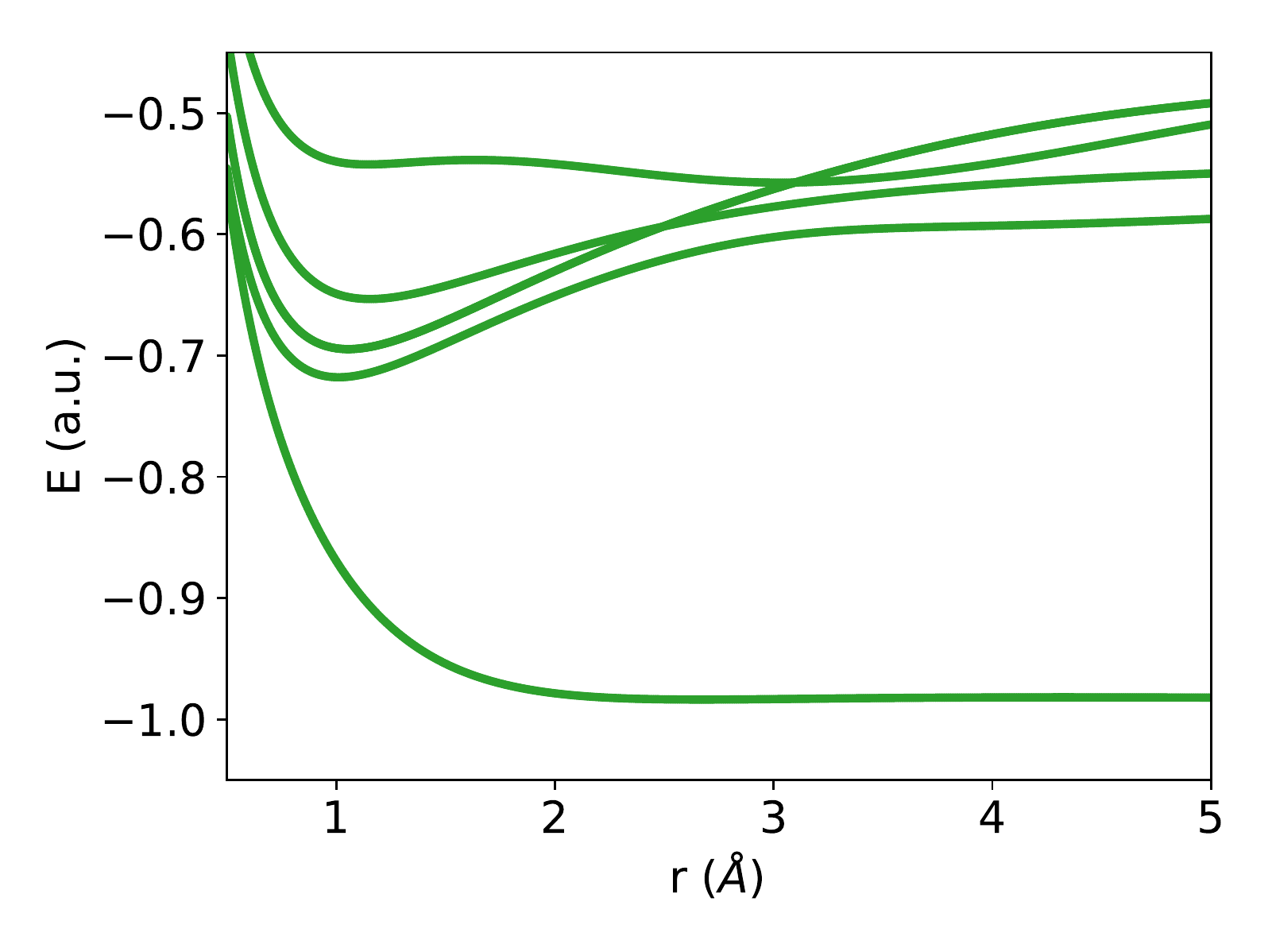}
        \caption{RCIS}
        \label{h2tz:Etriprcis}
    \end{subfigure}
        \begin{subfigure}[b]{0.48\textwidth}
        \centering
        \includegraphics[width=\linewidth]{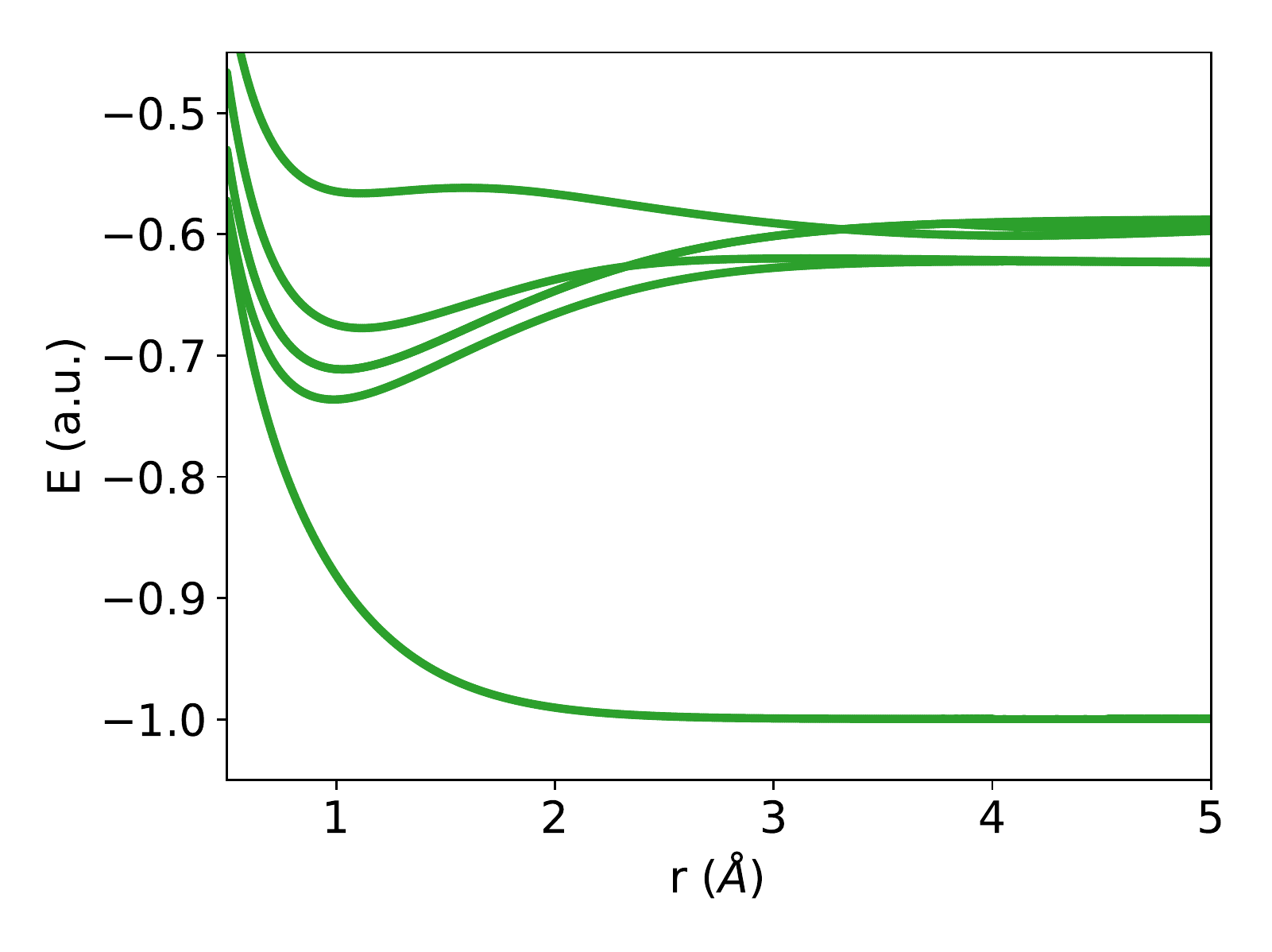}
        \caption{Exact}
        \label{h2tz:Etripexact}
    \end{subfigure}
    \caption{The CIS triplet excited state PESs of H$_2$ in the aug-cc-pVTZ basis compared to exact results. Small state crossing induced discontinuities might be present on the top surface. The UCIS $M_S=0$ triplet states are spin-contaminated beyond the CF point, but their analytical continuation is still followed to the dissociation limit.}
    \vspace*{-15pt}
    \label{h2tz:Trip}
\end{figure}

We next consider the behavior of the singlet and triplet excited states (and their continuations past the CF point) for the larger aug-cc-pVTZ basis, as shown in Figs. \ref{h2tz:Sing} and \ref{h2tz:Trip} respectively. Both singlet and triplet surfaces are significantly impacted by the CF point. Fig \ref{h2tz:Esing} shows that there is a sudden increase in energy for the singlet surfaces right beyond the CF point, with a clear first derivative discontinuity at $r=r_{\textrm{CF}}$. All depicted singlets develop some spin contamination as well (though any purely CT state would not have this issue). A comparison to exact surfaces in Fig \ref{h2tz:Esing} seems to suggest that all the affected states switch over to entirely new asymptotic regimes relative to their initial trajectory prior to the CF point, often accompanied by a dramatic change in the PES curvature. Spin polarization therefore seems to connect two distinct surfaces beyond the CF point, via the kink. 

Similar behavior is observed for the low lying $M_S=0$ triplet surfaces (Fig. \ref{h2tz:Etrip}), where shallow minima are effectively deepened relative to the dissociation limit via spin polarization. The $M_S=0$ T$_1$ state shows artifacts that resemble those seen in the minimal basis, slowly losing triplet character to become a CT state at intermediate separation. At long separations, it again spin polarizes to form an excited state localized on a single H atom with $\expt{S^2}=1$, which is lower in energy than the CT state. All character of the exact T$_1$ state (where both electrons are essentially on linear combinations of 1s orbitals) is consequently erased. In contrast, Fig \ref{fig:h2_tz_sfs} shows that the $M_S = \pm 1$ T$_1$ states remain qualitatively acceptable in larger basis sets and go to the correct dissociation limit. Higher energy triplets still have a kink at the CF point on account of spin polarization, and there is not much improvement in energy relative to $M_S=0$ subspace (cf. Fig \ref{h2tz:Etrip}), despite the states now being perfectly spin pure. The dramatic improvement in the quality of the T$_1$ state is nonetheless very promising, as it is often the principal actor in photodissociation.   

It is worth noting that the RCIS singlet and triplet surfaces (depicted in Figs. \ref{h2tz:Esingrcis} and \ref{h2tz:Etriprcis} respectively) are smooth, and appear to be mostly physical in comparison to the exact surfaces, despite being somewhat higher in energy (especially in the dissociation limit) due to missing correlation. It also appears that many excited states are slower to reach their asymptotic limits (relative to UCIS/FCI), as evidenced by relatively large slopes at even $5$ {\AA} separation. This could be a consequence
of CT character of the RHF reference being carried over to the excited states, as CT state energies asymoptotically decay as $r^{-1}$ (vs valence excitation energies, which decay exponentially to the asymptotic limit, like the fragment wave function overlap). This is however difficult to characterize for H$_2$ as there is no \textit{net} charge transfer, and only a two electron property (like a pair distribution function) would therefore be able to reveal whether the excited RCIS states have spurious CT character like the RHF reference. The T$_1$ state however asymptotes at a reasonable rate and appears to reach close to the correct dissociation limit of independent ground state H atoms (though is too high in energy by 0.47 eV). The qualitatively acceptable performance of the RCIS T$_1$ excited state however comes at the cost of a severely compromised RHF S$_0$ ground state, where spurious CT contributions drive it above the T$_1$ state by 0.21 a.u. (5.7 eV) at the dissociation limit!

\begin{figure}[h]
    \begin{subfigure}[b]{0.48\textwidth}
        \centering
        \includegraphics[width=\linewidth]{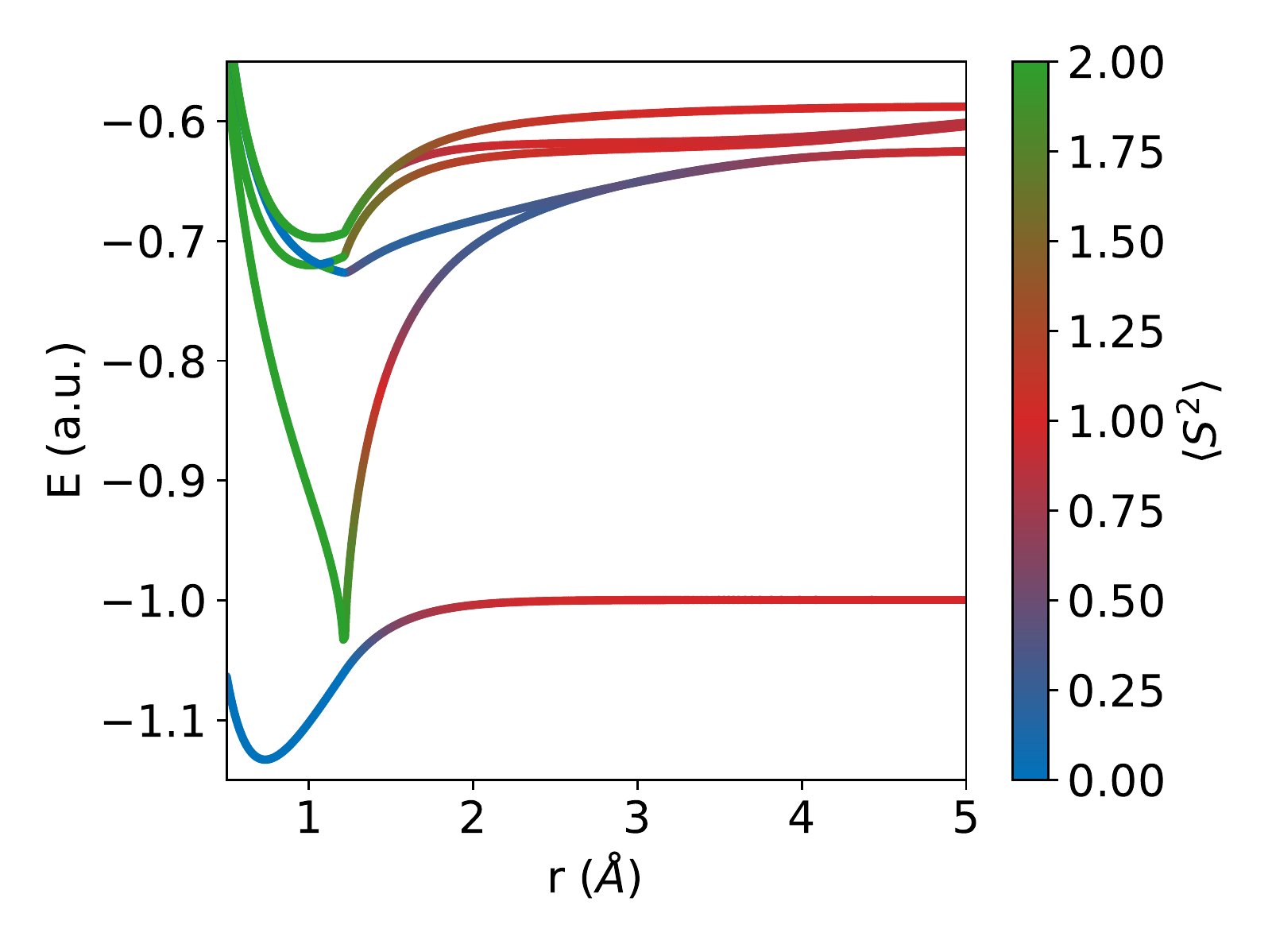}
        \subcaption{$M_S=0$ block.}
        \label{h2tz:tdhf0}
    \end{subfigure}
        \begin{subfigure}[b]{0.48\textwidth}
        \centering
        \includegraphics[width=\linewidth]{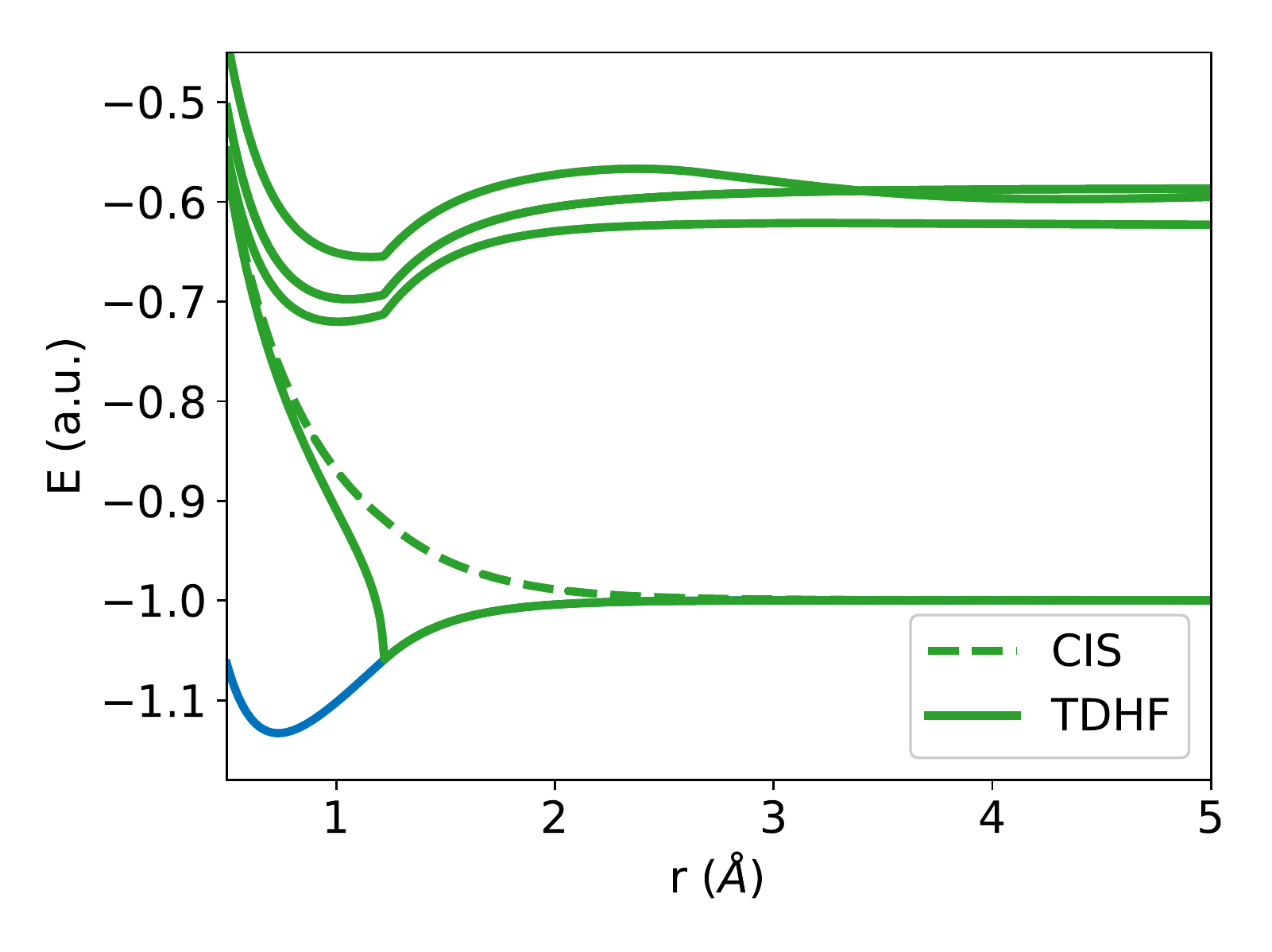}
        \subcaption{Spin-flipping block.}
        \label{h2tz:tdhf1}
    \end{subfigure}
    \caption{TDHF excited state PESs of H$_2$ in the aug-cc-pVTZ basis. The states have been labeled with the $\expt{S^2}$ of the corresponding CIS state. The $M_S=\pm 1$ CIS state has also been supplied as a reference. }
    \label{h2tz:tdhf}
\end{figure}

Moving on to TDHF, Fig \ref{h2tz:tdhf0} shows that the $M_S=0$  TDHF energy surfaces look similar to the CIS ones shown earlier in Figs \ref{h2tz:Esing} and \ref{h2tz:Etrip}. The quality of the T$_1$ surface is worse, as was the case for the minimal basis. Fig \ref{h2tz:tdhf1} shows that the $M_S=\pm 1$ T$_1$ TDHF solutions become degenerate with the UHF S$_0$ state past the CF point, as was the case with minimal basis as well. The $M_S=\pm 1$ higher excited states however are very similar to those obtained from CIS, including kinks at the CF point. Restricted orbitals offer no benefits for TDHF, as the instability in the RHF solution beyond the CF point causes the T$_1$ excitation energy to become imaginary (via Eqn \ref{eqnreform}). The dissociation limit S$_1$ excitation energy is also known to spuriously go to zero for H$_2$, and for all other symmetric bond dissociations\cite{giesbertz2008failure}.  Use of full TDHF (TDDFT) is therefore unlikely to lead to any qualitative improvements for excited state PESs around and beyond the CF point, relative to CIS (TDA). 

\begin{figure}[htb!]
\begin{minipage}{0.48\textwidth}
    \centering
    \includegraphics[width=\linewidth]{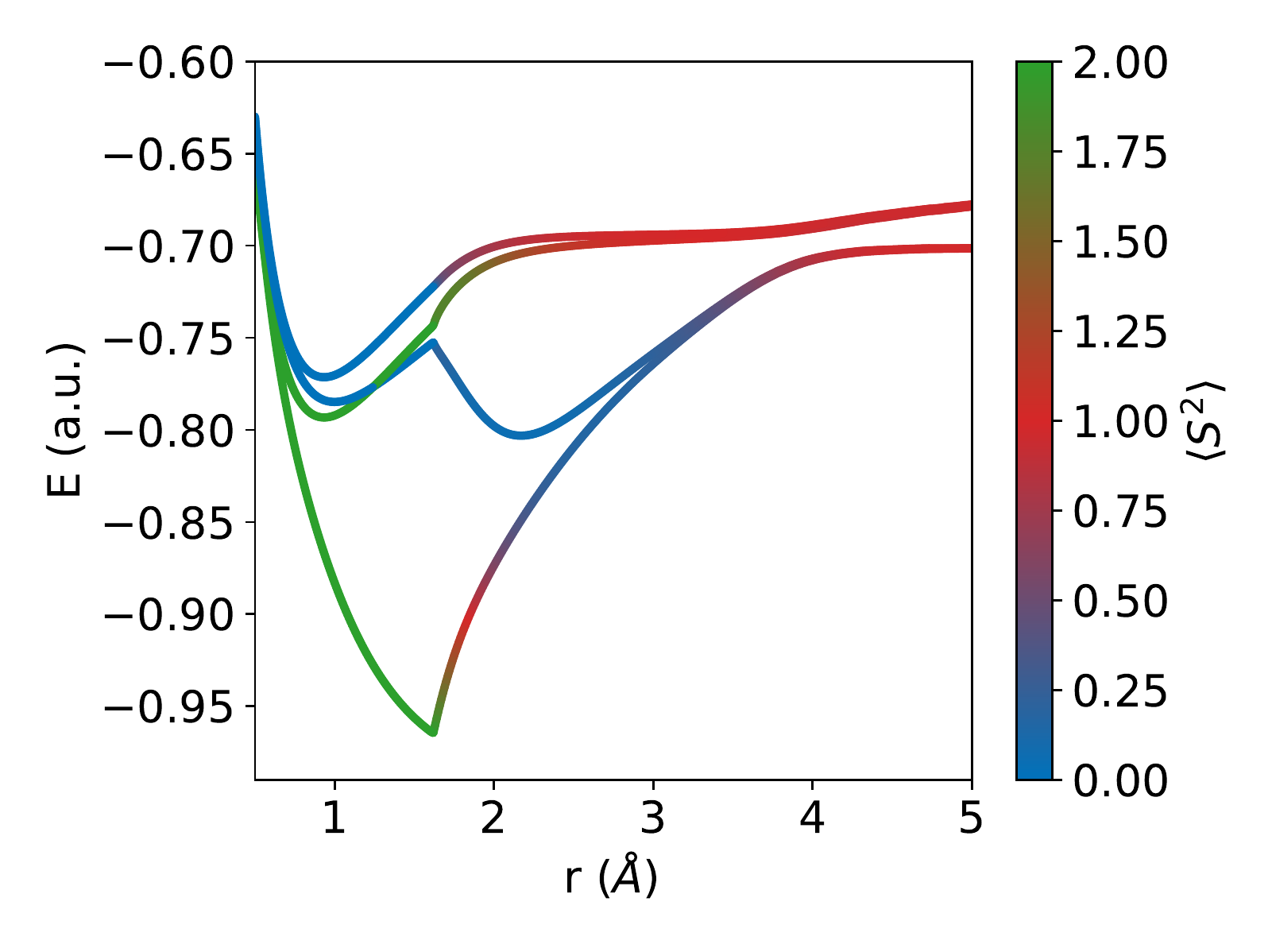}
    \subcaption{PBE}
    \label{fig:h2_tz_pbe}
\end{minipage}
\begin{minipage}{0.48\textwidth}
    \centering
    \includegraphics[width=\linewidth]{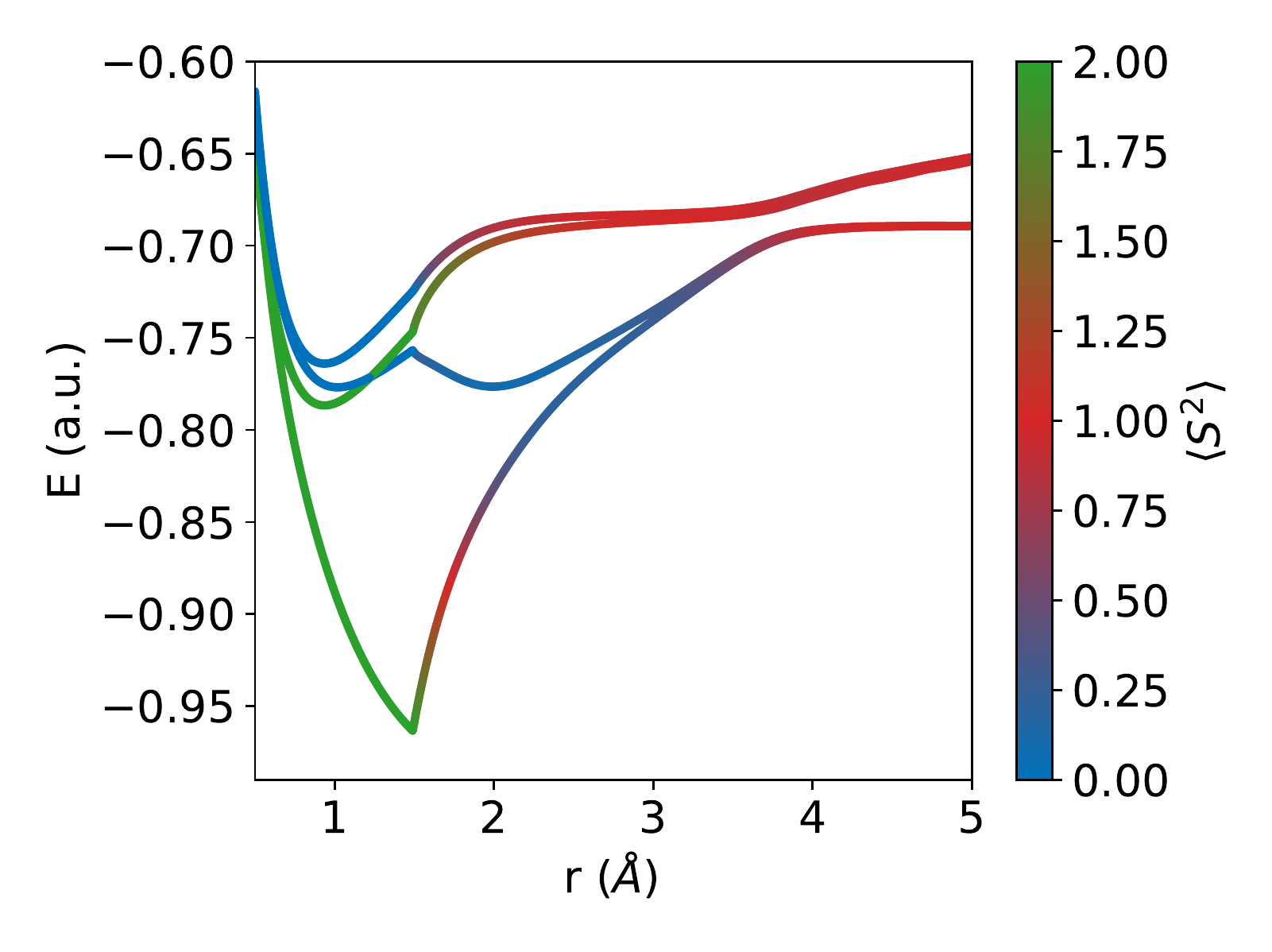}
    \subcaption{LRC-$\omega$PBEh}
    \label{fig:h2_tz_wpbe}
\end{minipage}
\caption{TDDFT/TDA (within the $M_S=0$ subspace) for stretched H$_2$/ aug-cc-pVTZ.}
\label{fig:h2_tz_dft}
\end{figure}

The observations discussed so far stem from the general formalism of TDDFT versus any specific issues with the HF functional. It therefore seems that CIS/TDHF results should be transferable to other density functionals, with minor adjustments. We demonstrate this by providing TDDFT/TDA PES for the local PBE\cite{PBE} and range separated hybrid LRC-$\omega$PBEh\cite{lrcwpbeh} functionals (within the $M_S=0$ subspace) in Fig. \ref{fig:h2_tz_dft}. We can see that the rapid increase beyond the CF point is somewhat less steep for non T$_1$  states in these cases, but this is solely on the account of the CF point occurring at longer $r$ due to presence of dynamic correlation. 

\begin{figure}[htb!]
    \begin{subfigure}[b]{0.48\textwidth}
        \centering
        \includegraphics[width=\linewidth]{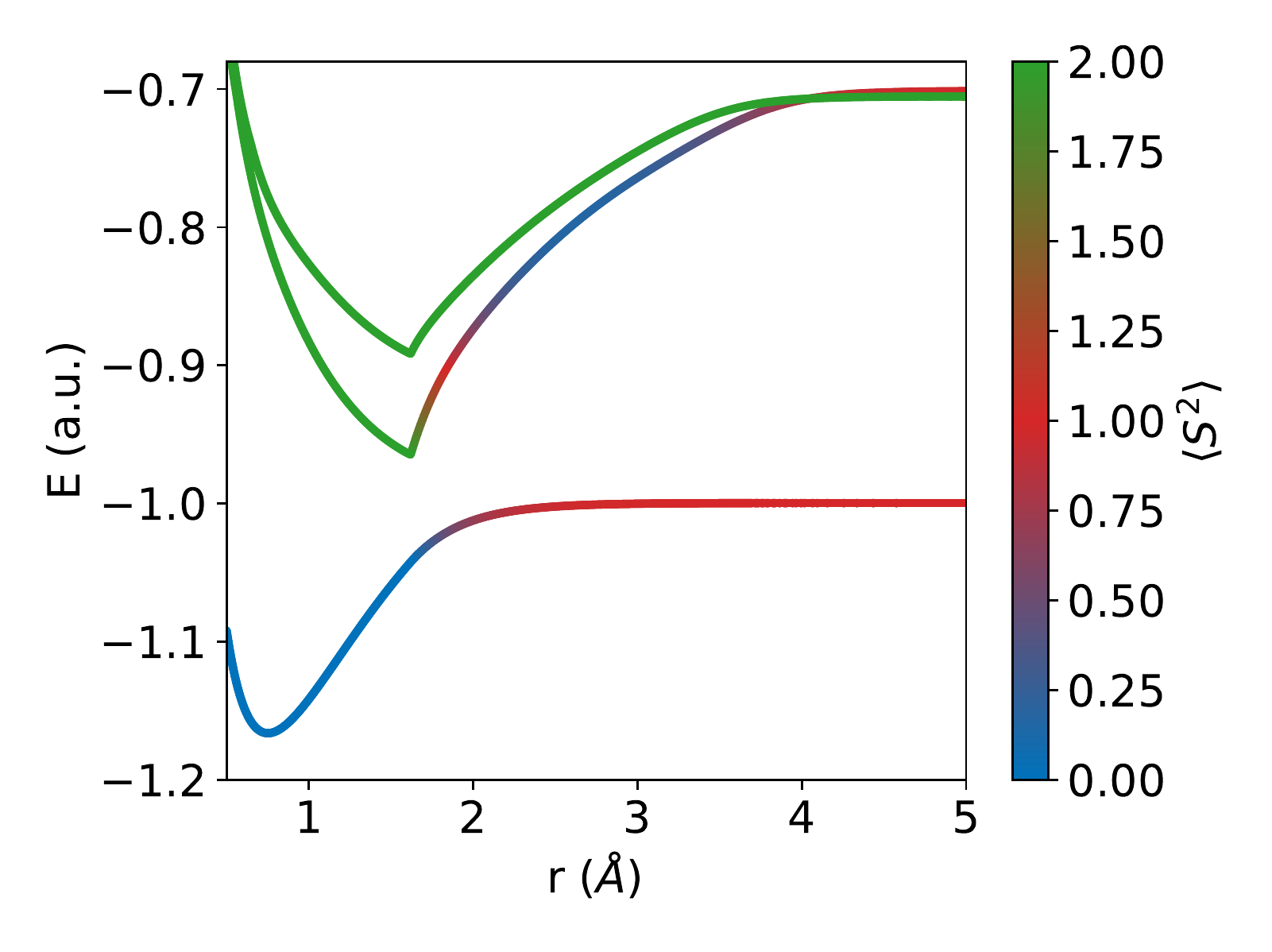}
        \caption{PBE}
        \label{h2tz:pbesfs}
    \end{subfigure}
    \begin{subfigure}[b]{0.48\textwidth}
        \centering
        \includegraphics[width=\linewidth]{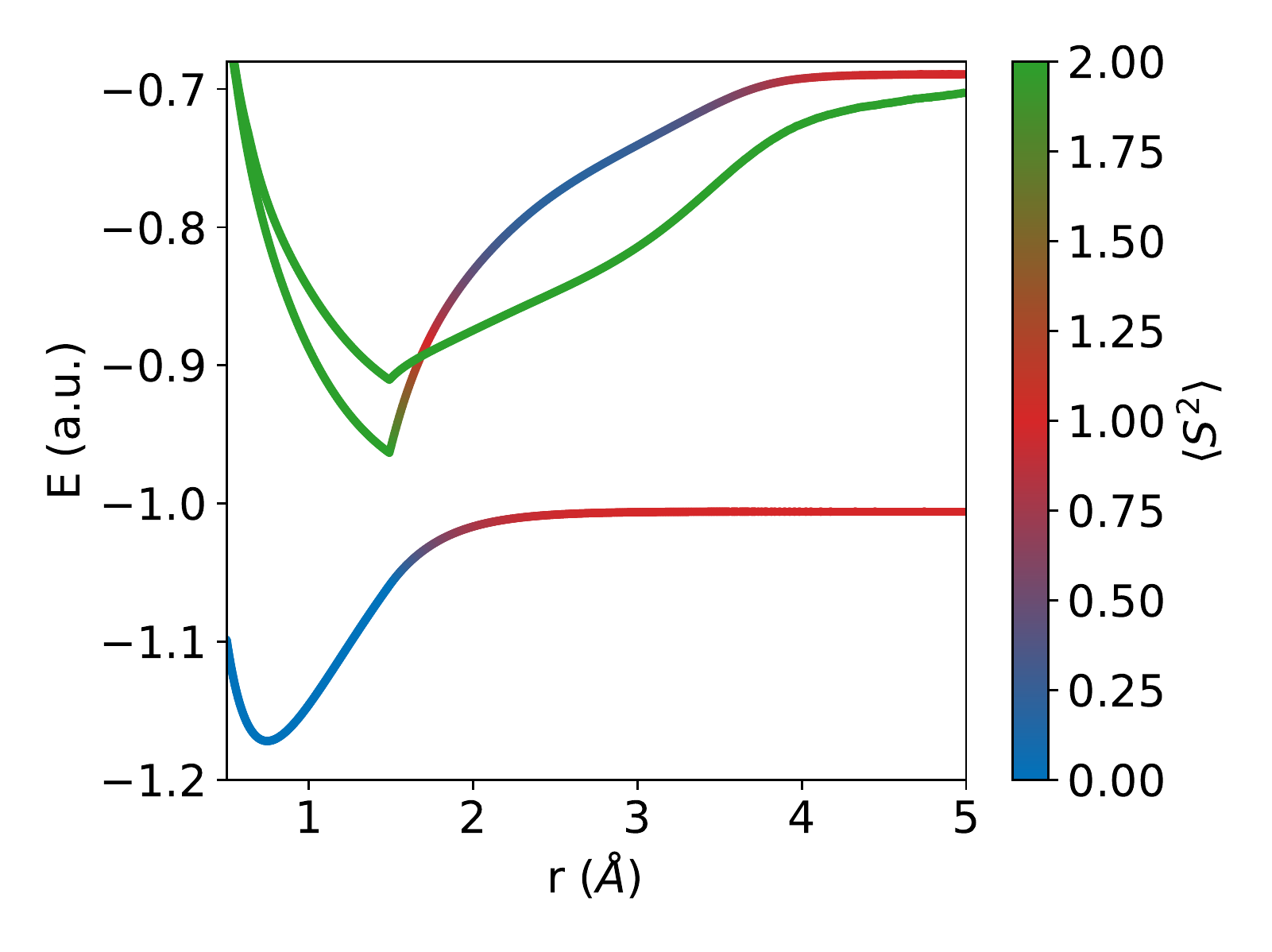}
        \caption{LRC-$\omega$PBEh}
        \label{h2tz:wpbehsfs}
    \end{subfigure}
        \begin{subfigure}[b]{0.48\textwidth}
        \centering
        \includegraphics[width=\linewidth]{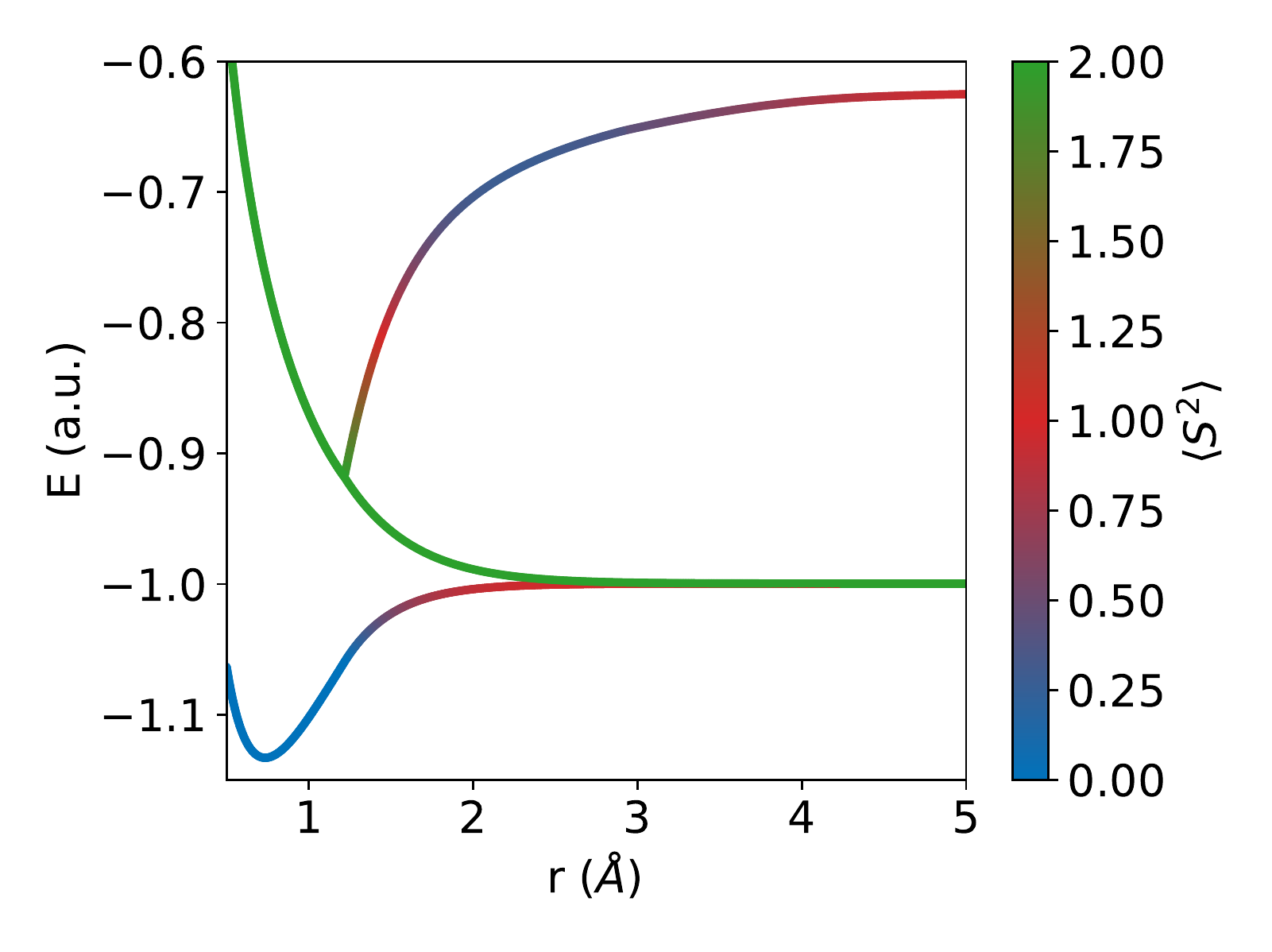}
        \caption{CIS}
        \label{h2tz:cis}
    \end{subfigure}
    \caption{T$_1$ surfaces within the $M_S=\pm 1$ and $M_S=0$ manifolds, as predicted by TDDFT/TDA. The $M_S=\pm 1$ states are spin pure (i.e. are purely green) while the $M_S=0$ state shows spin-polarization past the CF point. The ground UKS/UHF S$_0$ state is also supplied for comparison.}
    \label{h2tz:comb}
\end{figure}

The behavior for the spin-flipped block in TDDFT/TDA is  interesting, on account of the lack of local exchange-correlation contributions to the $f_{xc}$ kernel in the spin-flipped block. Fig. \ref{h2tz:comb} depicts the T$_1$ state obtained from the $M_S=\pm 1$ subspaces, along with the S$_0$ state and $M_S=0$ T$_1$, for PBE, LRC-$\omega$PBEh and HF/CIS. An important difference is evident prior to the onset of spin-polarization: the $M_S=\pm 1$ triplets predicted by the two density functionals are not degenerate with the $M_S=0$ triplet, as should be the case in exact quantum mechanics (and as is the case in CIS till the CF point). The lack of degeneracy between the different subspaces in TDDFT/TDA even prior to spin-polarization in the reference is due to the lack of the local exchange-correlation contributions to $f_{xc}$.

Beyond the CF point, the TDDFT/TDA $M_S=\pm 1$ T$_1$ states start mimicking the behavior of their $M_S=0$ counterpart, in sharp contrast to CIS. The $M_S=\pm 1$ T$_1$ surfaces display a sharp kink and rapid rise in energy to a dissociation limit that is non-degenerate with the S$_0$ state (as can be seen from Figs. \ref{h2tz:pbesfs} and \ref{h2tz:wpbehsfs}). The origin of this behavior can be understood from the form of Eqn \ref{Agen}. For a local functional like PBE, the $f_{xc}$ contribution is zero. Simultaneously, $i$ and $a$ have different spins (as do $j$ and $b$), voiding the $\braket{ij}{ab}$ term. The spin-flip block of $\mathbf{A}$ therefore becomes a purely diagonal matrix of orbital energy differences, and all excitation energies correspond to those diagonal elements. The non-degeneracy between the $\alpha$ HOMO and the $\beta$ LUMO (or vice versa) is nearly always guaranteed on account of the spin-polarized UKS potential they experience (even if the spatial orbitals are identical, as is the case for dissociation limit H$_2$). Consequently, the excitation energy cannot exactly become zero and the T$_1$ surface has to unphysically distort to accommodate this incorrect asymptotic behavior. The same general behavior applies to even hybrid functionals. Consequently, we conclude that the spin-flipped block yields physical T$_1$ surfaces only for CIS.

\section{Examples for other systems}
\begin{figure}[htb!]
\begin{minipage}{0.48\textwidth}
    \centering
    \includegraphics[width=\linewidth]{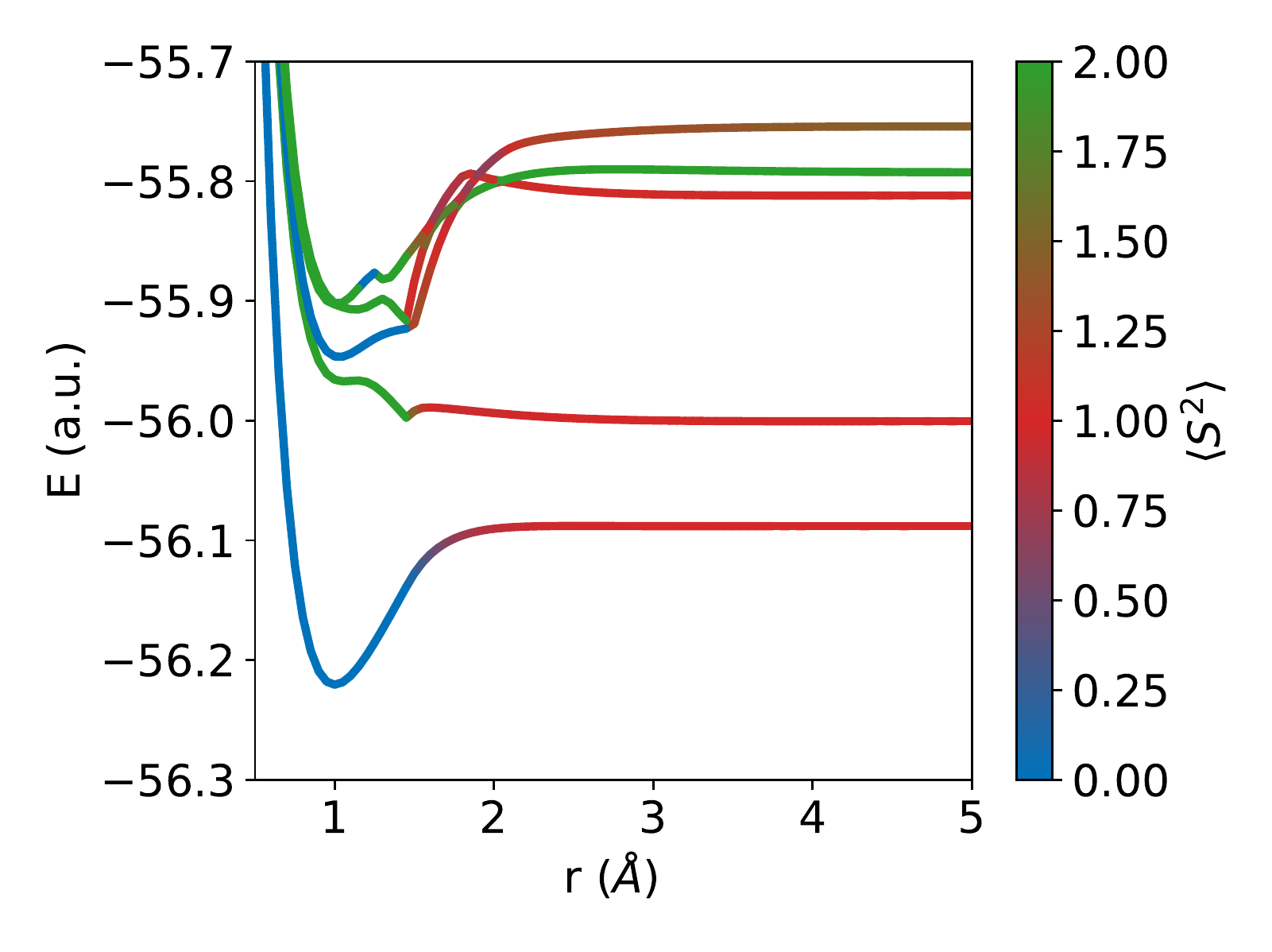}
    \subcaption{NH$_3$/aug-cc-pVTZ along N-H stretch.}
    \label{fig:nh3}
\end{minipage}
\begin{minipage}{0.48\textwidth}
    \centering
    \includegraphics[width=\linewidth]{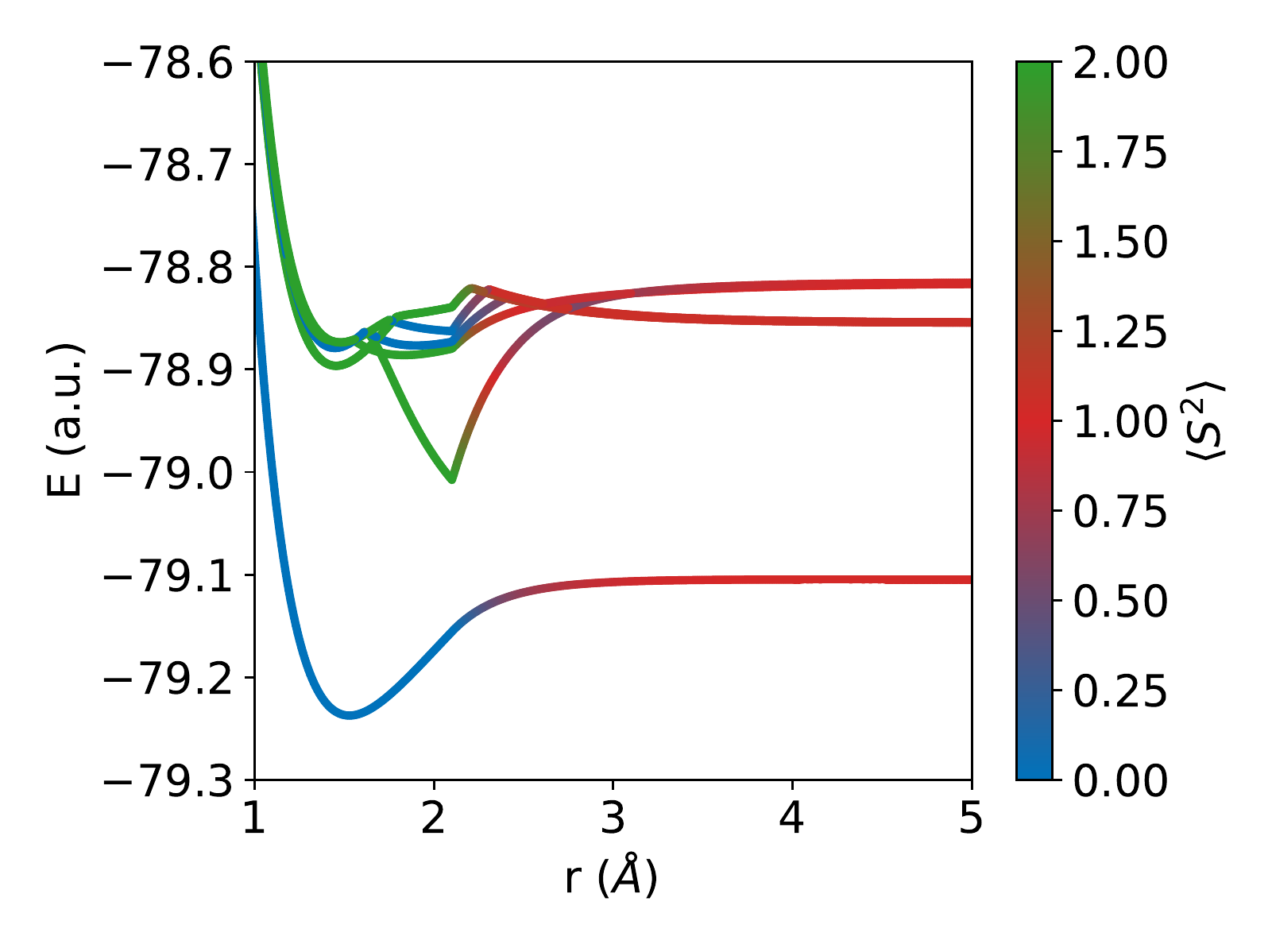}
    \subcaption{C$_2$H$_6$/aug-cc-pVDZ along C-C stretch.}
    \label{fig:ethane}
\end{minipage}
\caption{Ground and CIS excited states for larger species, within the M$_S$=0 subspace.}
\label{fig:cislarger}
\end{figure}

\subsection{$M_S=0$ subspace}

Fig. \ref{fig:cislarger} shows the behavior of $M_S=0$ UCIS for larger systems like the polar NH$_3$ and nonpolar ethane molecules. We see kinks appear at the CF point as well, along with dramatic jumps in many energy surfaces. The continuation of the T$_1$ state also does not go to the ground state dissociation limit. This indicates that our conclusions in the preceding subsection transfer to systems larger than H$_2$ as well, where more exact methods are no longer computationally affordable.

\subsection{$M_S=\pm 1$ subspaces}

CIS within the $M_S=\pm 1$ subspaces was effective in producing reasonable T$_1$ surfaces for H$_2$ (as can be seen from Fig \ref{fig:h2_tz_sfs}). This state of affairs however does not generalize to more complex systems, especially when polar bonds are involved. Let us now consider a hypothetical system where an A--B bond is stretched, with the unpaired $\alpha$ electron localizing on fragment A while the unpaired $\beta$ electron localizes on fragment B. To access the T$_1$ state with $M_S=1$, we would need to flip the unpaired $\beta$ electron  on B to an $\alpha$ virtual orbital. The resulting state however would have a B fragment with $M_S=\frac{1}{2}$ but orbitals optimized for $M_S=-\frac{1}{2}$ configuration (i.e the unrestricted fragment ground state). Other spin-flip excitations can mimic some orbital rotation effects, but aside from the trivial case of B being H, the resulting state is thus higher in energy than the unrestricted ground state that has B with $M_S=-\frac{1}{2}$ and self-consistent orbitals, leading to the T$_1$ state going to an incorrect dissociation limit. Essentially, while there exists no energetic penalty for flipping a spin to go from $M_S=\pm \frac{1}{2}$ to $M_S=\mp \frac{1}{2}$ in exact quantum mechanics, the same is not true for CIS. Table \ref{table:deltaab} supplies some representative values of the error in the CIS spin-flip energies (which we call $\Delta_{\alpha\beta}$) for going from $M_S=\pm \frac{1}{2}$ to $M_S=\mp \frac{1}{2}$, which are small if the species in question is H like (i.e., alkali metals, that have one valence electron atop a noble gas core) but can be substantial if the unpaired electron has other occupied orbitals close to it in energy (like CH$_3$ or NH$_2$).

\begin{table}
    \centering
    \begin{tabular}{| c | c |}
        \hline
        \textbf{Fragment} & \bm{$\Delta_{\alpha \beta}$}\\
        \hline
        Li         & 0.0014   \\
        Na         & 0.0022   \\
        CH$_3$    & 0.2504   \\
        C$_2$H$_5$ & 0.2724  \\
        SiH$_3$   & 0.0884  \\
        NH$_2$    & 0.2702 \\
        \hline
    \end{tabular}
    \caption{Degeneracy error (in eV) predicted by CIS for going from $M_S=\pm \frac{1}{2}$ to $M_S=\mp \frac{1}{2}$, for various radical fragments. }
    \label{table:deltaab}
\end{table}

This spurious degeneracy error can manifest itself in three different manners for systems more complex than H$_2$ (assuming the direction of spin polarization is consistent).

\begin{figure}[h]
    \centering
    \begin{minipage}[t]{0.48\textwidth}
    \centering
    \vspace{0pt}
    \includegraphics[width=\linewidth]{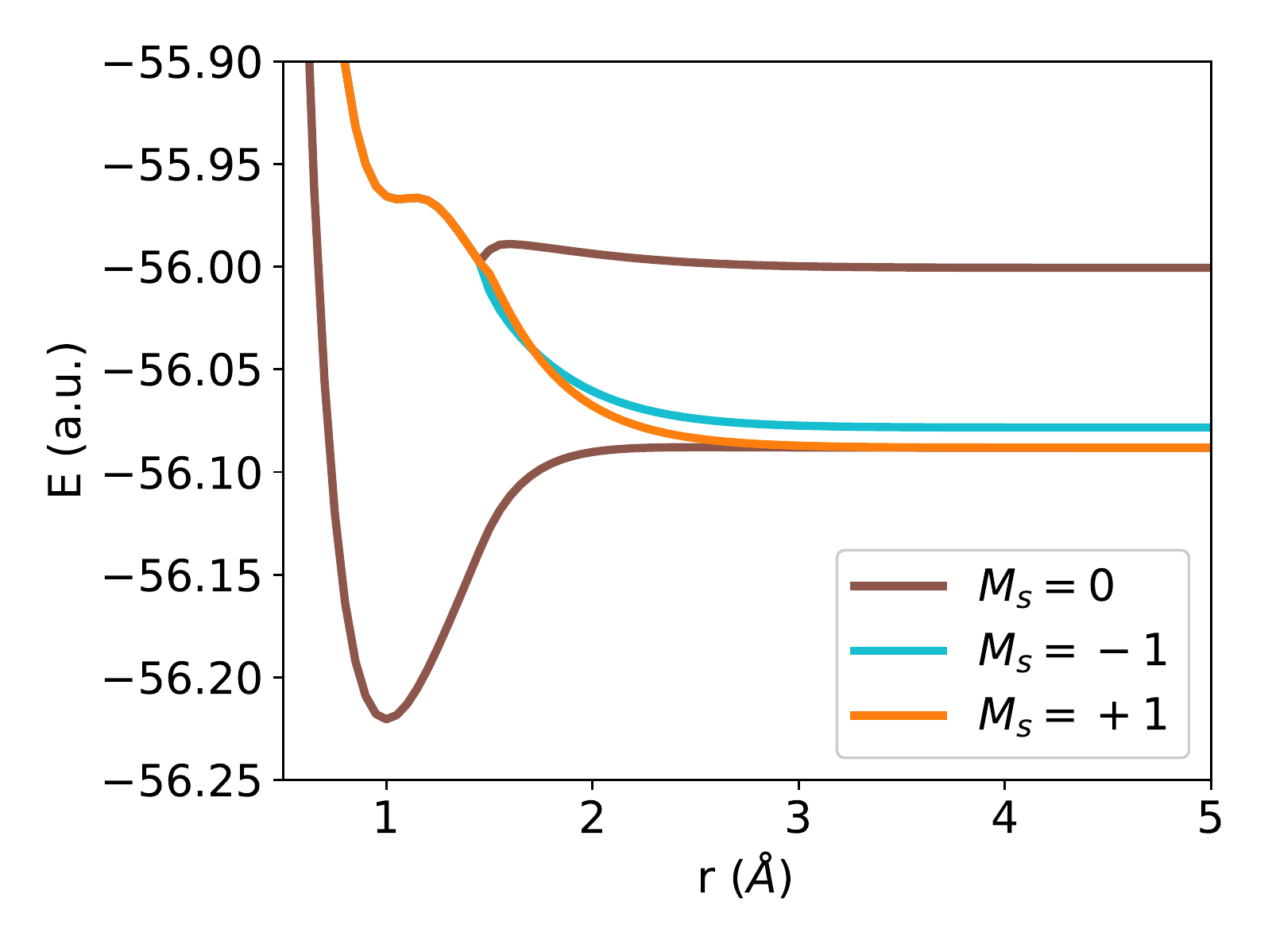}
    \subcaption{NH$_3$/aug-cc-pVTZ. The UHF ground state is also depicted, for comparison. }
    \label{fig:nh3_tz_sfs}
    \end{minipage}
    \begin{minipage}[t]{0.48\textwidth}
    \centering
    \vspace{0pt}
    \includegraphics[width=\linewidth]{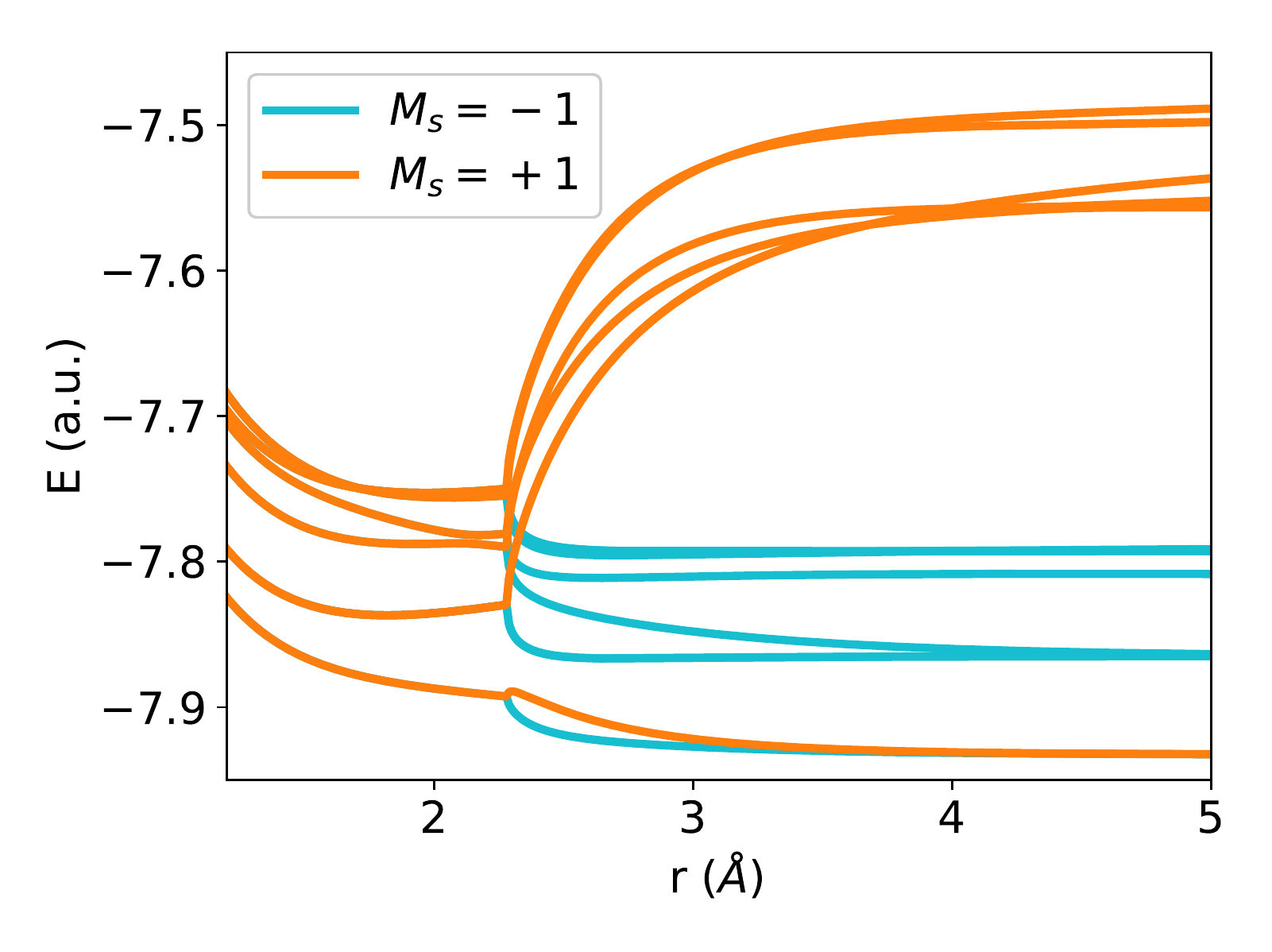}
    \subcaption{LiH/aug-cc-pVTZ. }
    \label{fig:lih_tz_sfs}
    \end{minipage}
    \caption{T$_1$ predicted by CIS for all three possible $M_S$.  The spin-polarization consistently placed the down spin on the H, so only the $M_S=1$ branch is asymptotically degenerate with the UHF ground state.}
\end{figure}

\begin{enumerate}
    \item For A-H bond dissociations, the subspace including spin inversion on H will have a T$_1$ state that becomes degenerate with the unrestricted S$_0$ state. The other branch (which includes spin inversion on A) has a T$_1$ state that remains above the correct dissociation limit by $\Delta_{\alpha\beta}^A$. An example of this can be seen in Fig \ref{fig:nh3_tz_sfs}, for NH$_3$.
    \item For A-A bond dissociations, both subspaces will yield degenerate solutions on account of symmetry. However, both T$_1$ solutions will be above S$_0$ by $\Delta_{\alpha\beta}^A$ in the asymptotic limit.
    \item For A-B bond dissociations where B$\ne$H, the two subspaces will yield different T$_1$ energies past the CF point, that remain distinct from the unrestricted ground state energy by $\Delta_{\alpha\beta}^A$ and $\Delta_{\alpha\beta}^B$ in the dissociation limit. 
\end{enumerate}

An additional feature for asymmetric bond dissociations is that the higher energy CIS solutions for the different $M_S$ are only degenerate up to the CF point, and subsequently move apart, going to quite distinct roots of different character in the dissociation limit. Consider the LiH molecule (as depicted in Fig \ref{fig:lih_tz_sfs}). If we place the unpaired $\alpha$ electron on Li during spin-polarization, then excitations to the $M_S=-1$ subspace contain only local excitations on Li but not on H. Similarly, excitations to the $M_S=1$ subspace capture local excitations on H , but not Li (barring very high energy core excitations). Similar considerations apply to CT transitions-it would be impossible to obtain a transfer of electron from H to Li in the $M_S=-1$ subspace, while only the core Li electron could be excited to H in the $M_S=1$. The individual $M_S=\pm 1$ subspaces therefore contain complementary information for asymmetric bond dissociations. 

\begin{figure}[h]
    \centering
    \begin{minipage}[t]{0.48\textwidth}
    \centering
    \vspace{0pt}
    \includegraphics[width=\linewidth]{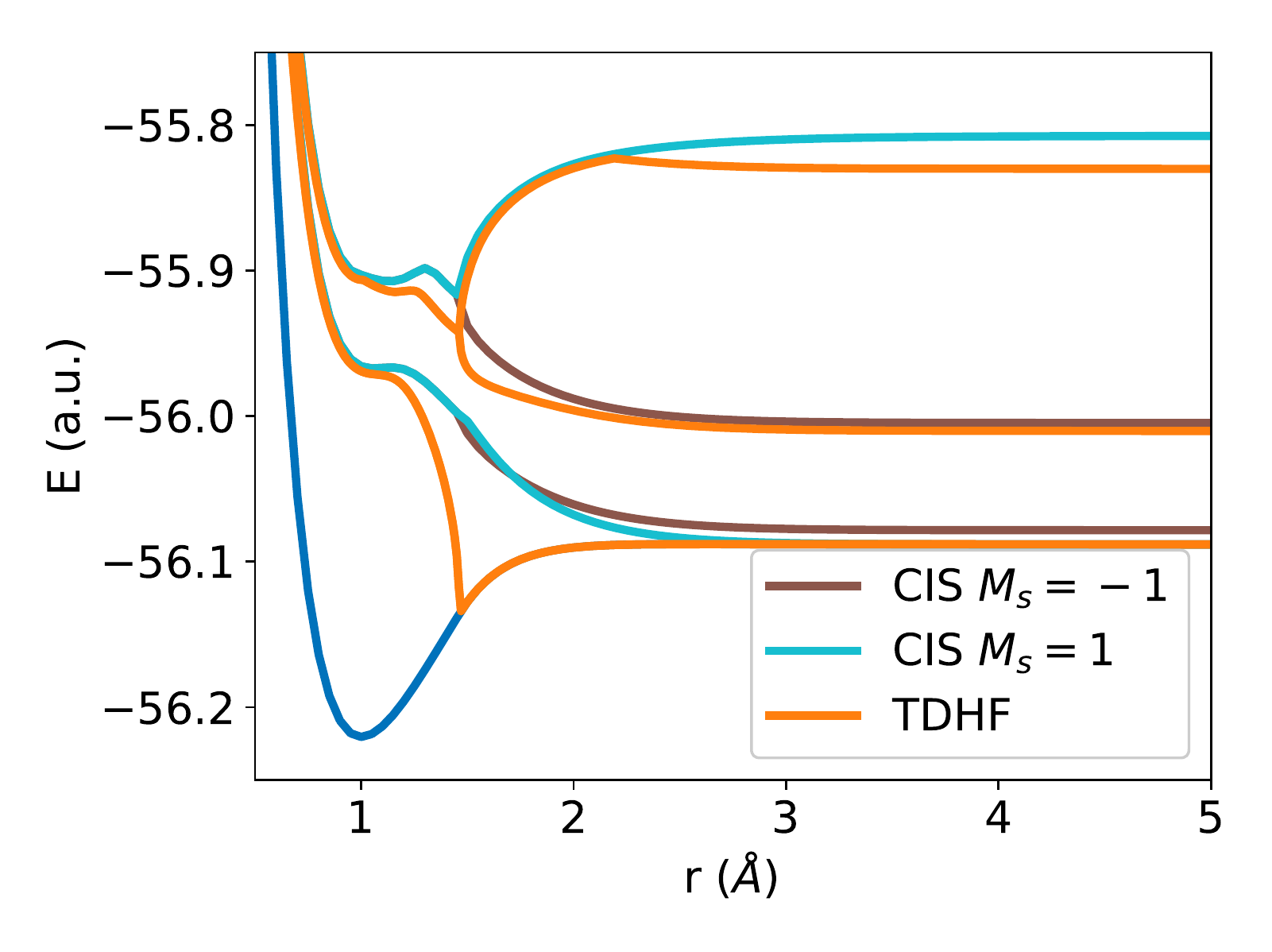}
    \subcaption{NH$_3$/aug-cc-pVTZ }
     \label{fig:nh3_tz_sfstdhf}
    \end{minipage}
    \begin{minipage}[t]{0.48\textwidth}
    \centering
    \vspace{0pt}
    \includegraphics[width=\linewidth]{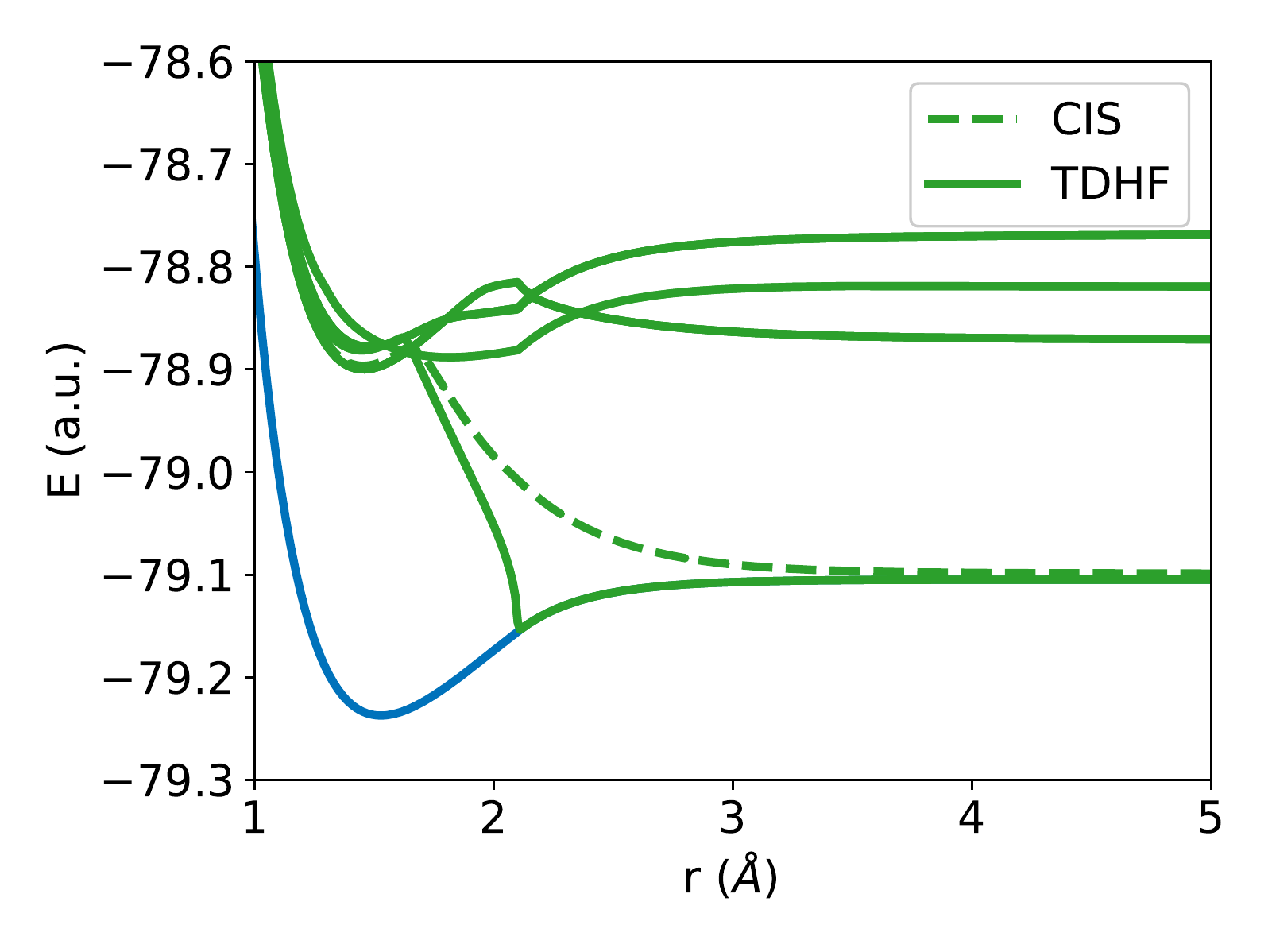}
        \subcaption{C$_2$H$_6$/aug-cc-pVDZ }
     \label{fig:ethane_augdz_sfstdhf}
    \end{minipage}
    \caption{Triplet states predicted by TDHF for the spin-flip block. Some CIS states are provided for comparison, as is the UHF ground state.}
     \label{fig:comp_sfstdhf}
\end{figure}

The spin-flip TDHF solutions remain degenerate with the $M_S=0$ solution till the CF point and then branch away. The T$_1$ states  merge with the S$_0$ ground state beyond the CF point (as can be seen from Fig. \ref{fig:comp_sfstdhf}), yielding zero excitation energy---just like minimal basis H$_2$. The higher excited states however resemble the corresponding CIS states, and show similar branching behavior (as can be seen from Fig \ref{fig:nh3_tz_sfstdhf}).

An interesting side consequence of the spin-flip T$_1$ states merging with the S$_0$ state is that the $\Delta_{\alpha\beta}$ induced degeneracy error in CIS is completely absent in TDHF. Indeed, calculations on open-shell radical fragments (with $M_S=\frac{1}{2}$) show that $\Delta_{\alpha\beta}=0$ in general. This can be viewed as a consequence explicit orbital response terms contained within the $\mathbf{B}$ matrix. An alternative interpretation draws upon the GHF stability argument given earlier in Sec. IV, by noting that direction of the spin on the open-shell fragment is itself arbitrary within GHF, and so there should not be any energetic cost for rotation to a different spin direction.

\begin{figure}[htb!]
\begin{minipage}{0.48\textwidth}
    \centering
    \includegraphics[width=\linewidth]{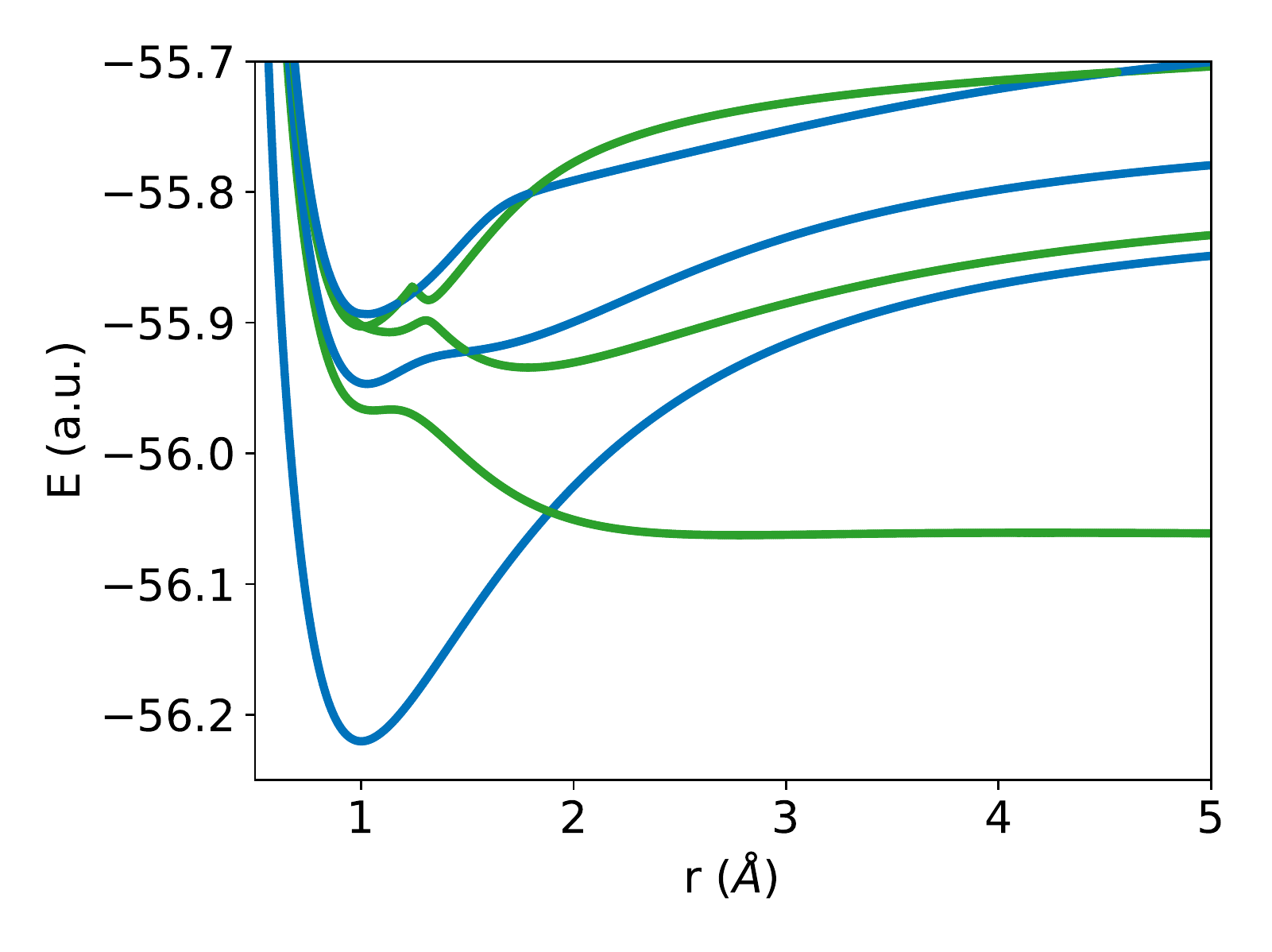}
    \subcaption{NH$_3$/aug-cc-pVTZ along N-H stretch.}
    \label{fig:nh3rest}
\end{minipage}
\begin{minipage}{0.48\textwidth}
    \centering
    \includegraphics[width=\linewidth]{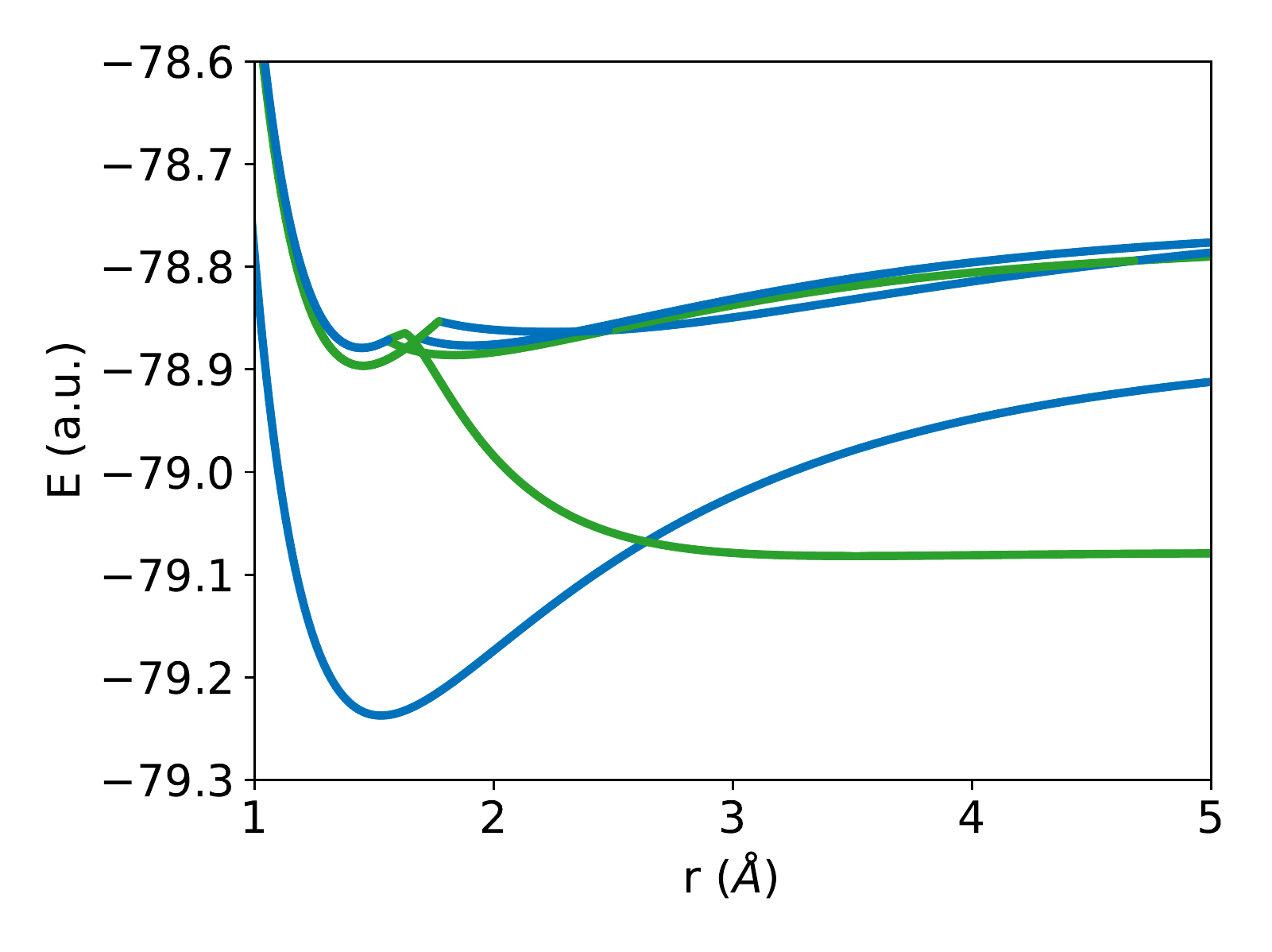}
    \subcaption{C$_2$H$_6$/aug-cc-pVDZ along C-C stretch.}
    \label{fig:ethanerest}
\end{minipage}
\caption{RHF S$_0$ and RCIS excited states for larger species.}
\label{fig:cislargerrest}
\end{figure}

\subsection{RCIS}
Figs \ref{fig:nh3rest} and \ref{fig:ethanerest} show RCIS excited state PESs for NH$_3$ and C$_2$H$_6$ respectively.  The T$_1$ surfaces in both cases appear to reach reasonable dissociation limits, which are nonetheless considerably below the S$_0$ RHF dissociation limit (on account of spurrious CT contributions in the latter). It is nonetheless worth noting that the asymptotic T$_1$ energies are above the dissociation limit of independent fragments by $\approx 0.5$ eV (a more complete listing given in Table  \ref{table:degenerr}). The T$_1$ surfaces are also not monotonically decreasing (as is physically expected) but rather have small local minima with depths of $\approx 6 $ kJ/mol or so relative to the dissociation limits.

\begin{table}
    \centering
    \begin{tabular}{| c | c |}
        \hline
        \textbf{Bond} & Asymptotic T$_1$ energy (in eV)\\
        \hline
        H---H & 0.47 (0.47)\\
        Li---H         & 0.60 (0.60)   \\
        Na---H        & 0.63 (0.63)   \\
        CH$_3$---CH$_3$ & 0.50 (0.67)\\
        C$_2$H$_5$---H & 0.78 (0.87)  \\
        SiH$_3$---H   & 0.54 (0.58)  \\
        NH$_2$---H    & 0.55 (0.62) \\
        \hline
    \end{tabular}
    \caption{Asymptotic T$_1$ energies (relative to ROHF independent fragments) predicted by RCIS for dissociation of A---B bonds. The T$_1$ energies relative to UHF fragments are given in parentheses. }
    \label{table:degenerr}
\end{table}

A comparison of Fig \ref{fig:cislargerrest}  with Fig \ref{fig:cislarger} also shows that at longer separations, the RCIS surfaces are a lot less flat relative to the UCIS ones. This is likely on account of spurious CT character in the excited states, which is confirmed by examining the dipole moments of RCIS excited states of NH$_3$ at 50 {\AA}. Aside from the T$_1$ state (which is nearly fully covalent), nine out of the ten lowest lying excited states have dipole moments in excess of 100 D, suggesting a fractional charge of $\approx 0.4$ on fragments. Similar behavior is seen for LiH/aug-cc-pVTZ at 50 {\AA} separation, indicating that this is not an unsual occurance for polar bonds. Interestingly, many excited state dipoles indicate charge transfer in the direction contrary to expectations based on electronegativity (i.e. Li$^{-\delta}$H$^{+\delta}$ vs the expected Li$^{+\delta}$H$^{-\delta}$), which is likely a consequence of CIS attempting to reverse the CT contamination in the RHF reference, but ultimately overcorrecting due lack of complete orbital response\cite{hait2018communication}. There is no net excited state dipole moment for low lying excited states of C$_2$H$_6$ because of the non-polar nature of the dissociating bond, but CT contributions are nonetheless present in both the reference and excited states, leading to incorrect asymptotic behavior of PESs. In general therefore, RCIS yields reasonable T$_1$ surfaces, but higher excited states often have substantial CT contamination on account of the RHF reference. 

\section{Conclusion} \label{conclusion}
In conclusion, we have characterized TDDFT excited states (as well as those predicted by the related TDHF and CIS methods) for single bond dissociation, concentrating on the region beyond the Coulson-Fischer point where the ground state spin polarizes. We find that spin-polarization in the stable UKS (UHF) state beyond the CF point leads to  disappearance of the initial $M_S=0$ T$_1$ state at long internuclear separations. This is a consequence of the unrestricted excited state corresponding to the $M_S=0$ T$_1$ having opposite spin-polarization than the ground state and is subsequently a double excitation away, despite being low lying in energy (and formally degenerate with the ground state at the dissociation limit). Kinks  at the CF point are observed in many other excited state surfaces, along with spuriously elevated dissociation limits and unphysical curvature. It in fact appears that the spin-polarization zone just beyond the CF point connects what should naturally be two different surfaces via a a rapidly increasing concave segment.

In CIS, triplet solutions solutions within the spin-flip $M_S=\pm 1$ subspaces are non-degenerate with the standard $M_S=0$ subspace beyond the CF point. The $M_S=\pm 1$ T$_1$ states roughly reach the correct dissociation limit for T$_1$. They are  however non-degenerate past the CF point (or do not become exactly asymptotically degenerate with the ground S$_0$ state) due to insufficient orbital relaxation effects in CIS, for most single bond dissociation processes. The TDHF spin-flipped T$_1$ states are exactly degenerate with the UHF S$_0$ ground state beyond the CF point. These spin-flipping subspaces exhibit less desirable behavior when a functional with any local exchange-correlation contribution is employed. The shapes of the $M_S=\pm 1$ T$_1$ state surfaces predicted by typical TDDFT/TDA  resemble the unphysical $M_S=0$ T$_1$ surface.

We note that restricted CIS yields reasonable T$_1$ surfaces, despite the incorrect ground state dissociation limit solution. Of course, full TDDFT (or TDHF) on RHF/RKS solutions beyond the CF point is unwise on account of unphysical complex (or vanishing) excitation energies. Even for CIS, it must be noted that higher energy RCIS excited states, while smooth in contrast to their UCIS counterparts, tend to have significant amounts of spurious CT contamination, suggesting this approach has limited applicability and should be viewed with caution. If the spurious ionic terms in the ground state can be reduced, such as via approximate coupled cluster methods, the resulting restricted excited states will be more useful. Alternatively, some type of non-orthogonal CI\cite{thom2009hartree,sundstrom2014non} could be employed to make the unrestricted methods useful, and the recently proposed holomorphic HF extensions\cite{burton2015holomorphic} look promising in this regard.

\section*{Computational Details}
All calculations were performed with the Q-Chem 5.2 \cite{QCHEM4} package. 	Local exchange-correlation integrals were calculated over
a radial grid with 99 points and an angular Lebedev grid with 590 points for all atoms. All internal coordinates other than the stretch of the dissociating bond were held frozen at equilibrium configuration for polyatomic species (e.g. CH$_3$---CH$_3$ dissociates into unrelaxed trigonal pyramidal CH$_3$ radicals instead of relaxed, trigonal planar CH$_3$ radicals), for simplicity. 
\section*{Conflicts of Interest}
There are no conflicts of interest to declare.

\section*{Acknowledgment} 
	This research was supported by the Director, Office of Science, Office of Basic Energy Sciences, of the U.S. Department of Energy under Contract No. DE-AC02-05CH11231. 

\appendix
\section{TDHF solutions to the SF block for minimal basis H$_2$}
It is possible to analytically show that the TDHF excitation energies for the spin-flipping $M_S=\pm 1$ block is zero for minimal basis H$_2$. We follow the treatment in Ref \onlinecite{szabo2012modern} and denote the RHF spatial orbitals as $\phi_1$ and $\phi_2$ (which are normalized symmetric and antisymmetric combinations of the atomic orbitals, respectively). These have been referred to as $\ket{\sigma}$ and $\ket{\sigma^*}$ earlier in the text, but have been relabelled for consistency with Ref \onlinecite{szabo2012modern}. The occupied $\alpha$ and $\beta$ spatial orbitals are consequently:
\begin{align}
    \phi_1^\alpha&=\phi_1\cos\theta+\phi_2\sin\theta\\
    \phi_1^\beta&=\phi_1\cos\theta-\phi_2\sin\theta
\end{align}
and the virtuals are similarly:
\begin{align}
    \phi_2^\alpha&=\phi_2\cos\theta-\phi_1\sin\theta\\
    \phi_2^\beta&=\phi_2\cos\theta+\phi_1\sin\theta
\end{align}
$\theta=0$ before the CF point (yielding the RHF solution) and goes to $\dfrac{\pi}{4}$ in the dissociation limit. 

Since there is only one occupied and one virtual orbital in both spin subspaces, the $\mathbf{A}$ and $\mathbf{B}$ matrices in the spin flip block are simply scalars. Mathematically, it implies:
\begin{align}
 \mathbf{A}_{\alpha\beta,\alpha\beta}&=\left(\epsilon^\beta_2-\epsilon^\alpha_1\right)+\braket{\alpha_1\alpha_1}{\beta_2\beta_2}-\braket{\alpha_1\beta_2}{\alpha_1\beta_2}\\
 \mathbf{A}_{\beta\alpha,\beta\alpha}&=\mathbf{A}_{\alpha\beta,\alpha\beta}\mbox{ (From spatial symmetry)}\\
  \mathbf{B}_{\alpha\beta,\beta\alpha}&=\braket{\alpha_1\alpha_2}{\beta_2\beta_1}-\braket{\alpha_1\beta_2}{\alpha_2\beta_1}=\mathbf{B}_{\beta\alpha,\alpha\beta}
\end{align}
However, the two electron integrals $\braket{\alpha_1\alpha_1}{\beta_2\beta_2}$ and $\braket{\alpha_1\alpha_2}{\beta_2\beta_1}$ are zero, as the electron-repulsion term of the matrix element has no spin-component, and so the spin parts integrate to zero. 

Let us furthermore denote matrix elements of the Hamiltonian in the RHF basis (as in Ref \onlinecite{szabo2012modern}), which gives us one electron matrix elements $h_{11}$ and $h_{22}$ and two electron matrix elements $\braket{11}{11}=J_{11}$ (self-repulsion in orbital $\phi_1$), $\braket{22}{22}=J_{22}$ (self-repulsion in orbital $\phi_2$), $\braket{12}{12}=J_{12}$ (repulsion between orbitals $\phi_1$ and $\phi_2$) and $\braket{12}{21}=K_{12}$ (exchange interaction between orbitals $\phi_1$ and $\phi_2$). Other terms like $h_{12}$ or $\braket{11}{12}$ cancel out during the simplification and do not enter the picture. 

With this notation, we have:
\begin{align}
    \mathbf{B}_{\alpha\beta,\beta\alpha}&=\left(J_{11}+J_{22}-2J_{12}\right)\cos^2\theta\sin^2\theta-\cos^22\theta K_{12}\\
    \mathbf{A}_{\alpha\beta,\alpha\beta}&=\cos2\theta\left(h_{22}-h_{11}\right)-\cos^4\theta J_{11}-\sin^4\theta J_{22}+\left(\cos^4\theta+\sin^4\theta\right)J_{12}-\cos^22\theta K_{12}\\
    &=\cos2\theta\left(h_{22}-h_{11}\right)-\cos^2\theta\cos2\theta J_{11}-\cos^2\theta\sin^2\theta J_{11}+\sin^2\theta\cos2\theta J_{22}-\sin^2\theta\cos^2\theta J_{22}\notag\\&+\cos^2 2\theta J_{12}+2\sin^2\theta\cos^2\theta J_{12}-\cos^22\theta K_{12}\\
    &=\cos2\theta\left(h_{22}-h_{11}-\cos^2\theta J_{11}+\sin^2\theta J_{22}+\cos 2\theta J_{12}-2\cos2\theta K_{12}\right)\notag\\&-\cos^2\theta\sin^2\theta J_{11}-\sin^2\theta\cos^2\theta J_{22}+2\sin^2\theta\cos^2\theta J_{12}+\cos^22\theta K_{12}
\end{align}

However, the spin-polarized UHF solution satisfies $h_{22}-h_{11}-\cos^2\theta J_{11}+\sin^2\theta J_{22}+\cos 2\theta J_{12}-2\cos2\theta K_{12}=0$ (Eqn. 3.374 in Ref \onlinecite{szabo2012modern}), leading to:
\begin{align}
    \mathbf{A}_{\alpha\beta,\alpha\beta}&=-\cos^2\theta\sin^2\theta J_{11}-\sin^2\theta\cos^2\theta J_{22}+2\sin^2\theta\cos^2\theta J_{12}+\cos^22\theta K_{12}=-\mathbf{B}_{\alpha\beta,\beta\alpha}
\end{align}

The eigenvalues of $\mathbf{A}_{\alpha\beta,\alpha\beta}$ are the CIS excitations energies $\omega_{CIS}$ within the SF block. Consequently, Eqn. \ref{plus1} simplifies to:
\begin{align}
    \begin{pmatrix} \omega_{CIS}&-\omega_{CIS}\\-\omega_{CIS} &\omega_{CIS}\end{pmatrix}\begin{pmatrix} X \\ Y\end{pmatrix} &=\omega_{TDHF}
    \begin{pmatrix} 1 & 0\\ 0 & -1\end{pmatrix}\begin{pmatrix} X \\ Y\end{pmatrix}\\
    \implies  \begin{pmatrix} \omega_{CIS}&-\omega_{CIS}\\\omega_{CIS} &-\omega_{CIS}\end{pmatrix}\begin{pmatrix} X \\ Y\end{pmatrix} &=\omega_{TDHF}
    \begin{pmatrix} X \\ Y\end{pmatrix}
\end{align}
whose only eigenvalue is $0$, corresponding to the $\begin{pmatrix} 1 \\ 1\end{pmatrix}$ eigenvector. Therefore, the TDHF excitation energies within the SF block are zero beyond the CF point, which is consistent with numerical observation.

\bibliography{references}

\end{document}